\documentclass[journal]{IEEEtran}
\usepackage{amsmath}
\usepackage{cite}
\usepackage{amsfonts,amssymb}
\usepackage{graphicx}
\usepackage{subfigure}
\usepackage{multirow}
\usepackage{enumerate}
\usepackage[shortlabels]{enumitem}
\usepackage[T1]{fontenc}
\usepackage{stfloats}
\usepackage{booktabs}
\usepackage[noend]{algorithmic}
\usepackage{tikz}
\usepackage[subnum]{cases}
\usepackage{algorithm}
\usepackage{array}
\usepackage{color}
\usepackage{colortbl}
\usepackage{mathtools}
\usepackage[keeplastbox]{flushend}
\usepackage{xcolor}
\usepackage{threeparttable}

\usetikzlibrary{trees}

\ifCLASSINFOpdf
\else
\fi
\algsetup{indent=2em}

\floatname{algorithm}{\bf{Summary of the proposed algorithm}}

\begin{document}
%
\title{Joint Blind Identification of the Number of Transmit Antennas and MIMO Schemes Using Gerschgorin Radii and FNN}
%
%
%

\author{
        Mingjun Gao, \IEEEmembership{Student Member,~IEEE,}
        Yongzhao Li, \IEEEmembership{Senior Member,~IEEE,}\\
        Octavia A. Dobre, \IEEEmembership{Senior Member,~IEEE,} and
        Naofal Al-Dhahir, \IEEEmembership{Fellow,~IEEE} 
}

\maketitle

\begin{abstract}
Blind enumeration of the number of transmit antennas and blind identification of multiple-input multiple-output (MIMO) schemes are two pivotal steps in MIMO signal identification for both military and commercial applications. Conventional approaches treat them as two independent problems, namely the source number enumeration and the presence detection of space-time redundancy, respectively. In this paper, we develop a joint blind identification algorithm to determine the number of transmit antennas and MIMO scheme simultaneously. By restructuring the received signals, we derive three subspace-rank features based on the signal subspace-rank to determine the number of transmit antennas and identify space-time redundancy. Then, a Gerschgorin radii-based method and a feed-forward neural network are employed to calculate these three features, and a minimal weighted norm-1 distance metric is utilized for decision making. In particular, our approach can identify additional MIMO schemes, which most previous works have not considered, and is compatible with both single-carrier and orthogonal frequency division multiplexing (OFDM) systems. Simulation results verify the viability of our proposed approach for single-carrier and OFDM systems and demonstrate its favorable identification performance for a short observation period with acceptable complexity.
\end{abstract}


%
\IEEEpeerreviewmaketitle

\section{Introduction}
%
%
%
%

\IEEEPARstart{B}{lind} identification of information signals' parameters of a transmitter from received signals has important applications both in military and civilian communication systems. In the context of military applications, this parametric knowledge can help an attacker to carry out electronic warfare operations, such as surveillance and jamming signal selection. Moreover, blind identification has found wide applications in civilian reconfigurable systems including software-defined and cognitive radios\cite{Survey_Signal_Identification}. Multiple-input multiple-output (MIMO) and orthogonal frequency division multiplexing (OFDM) technologies are adopted in cellular and WiFi standards because they achieve high spectral efficiency. Different from the identification of single-antenna systems, the blind identification of MIMO or MIMO-OFDM signals requires the enumeration of the number of transmit antennas\cite{AIC_MDL,eigenvector_nt_est,PET,WME,HOM_TD_Nt_est} and identification of MIMO schemes\cite{Likelihood_Based,correlator_function,higher_order_cyclic,blind_recognition_STBC,Hierarchical_STBC, Fourth_order_TC,Second_Order_cyclic,K_S_test,Classify_STBC_Over_FS,STBC_cyclic_2015_ICC,Blind_MIMO_OFDM,Blind_MIMO_OFDM_SM_AL,Identification_SM_AL_OFDM_cyclic,blind_SFBC,My_paper_Globecom,My_paper_TWC2,My_paper_TVT} for single-carrier or OFDM systems.

The identification of MIMO schemes is the process of classifying the spatial multiplexing (SM) or transmit diversity (TD) codes, namely space-time block codes (STBC), which utilize space-time redundancy to reduce the error rate. Previous works on the identification of MIMO schemes include \cite{Likelihood_Based,correlator_function,higher_order_cyclic,blind_recognition_STBC,Hierarchical_STBC, Fourth_order_TC,Second_Order_cyclic,K_S_test,Classify_STBC_Over_FS,STBC_cyclic_2015_ICC} for single-carrier systems and \cite{Blind_MIMO_OFDM,Blind_MIMO_OFDM_SM_AL,Identification_SM_AL_OFDM_cyclic,blind_SFBC,My_paper_Globecom,My_paper_TWC2,My_paper_TVT} for OFDM systems. Regarding the identification of MIMO schemes for single-carrier systems, previous works follow either likelihood-based \cite{Likelihood_Based} or feature-based \cite{correlator_function,higher_order_cyclic,blind_recognition_STBC,Hierarchical_STBC, Fourth_order_TC,Second_Order_cyclic,K_S_test,Classify_STBC_Over_FS,STBC_cyclic_2015_ICC} methods. The former relies on the likelihood function of the received signals to quantify the space-time redundancy and classify different STBCs. The latter detects the presence of the space-time redundancy at some specific time-lag locations based on the features of signal statistics or cyclic statistics; however, it can only identify a small number of STBC types due to identical features for several SFBCs. Although \cite{STBC_cyclic_2015_ICC} can identify 11 types of STBCs utilizing the feature of second-order cyclostationary statistics, it requires the number of transmit antennas and channel coefficients as \textit{a priori} information. 

As for OFDM systems, there are two main approaches to combine the TD codes with OFDM signals. The first approach is STBC-OFDM, where the diversity STBC is implemented over consecutive OFDM symbol intervals. STBC-OFDM has been adopted in several indoor MIMO wireless standards, such as WiFi\cite{IEEE802_11,STBC_on_80211}, owing to its excellent performance. Another TD coding scheme used in OFDM systems is space-frequency block coding (SFBC)-OFDM, where the diversity SFBC is employed over consecutive subcarriers of an OFDM symbol. Several cellular wireless standards supporting high mobility, such as LTE\cite{sesia2009lte} and WiMAX \cite{IEEE802_16}, favor SFBC-OFDM over STBC-OFDM because of its superior performance in a high mobility environment \cite{ICI_affects_OSTBC}. The identification approaches for single-carrier systems fail to identify MIMO schemes of OFDM systems under frequency-selective fading channels due to multipath effects. Previous works on the identification of STBC-OFDM systems include \cite{Blind_MIMO_OFDM,Blind_MIMO_OFDM_SM_AL,Identification_SM_AL_OFDM_cyclic}, which detect the presence of space-time redundancy based on the peaks of the cross-correlation functions between two receive antennas in the time-domain. Specifically, \cite{Blind_MIMO_OFDM,Blind_MIMO_OFDM_SM_AL} use different cross-correlation functions, while \cite{Identification_SM_AL_OFDM_cyclic} employs a cyclic cross-correlation function with a specific time-lag during adjacent OFDM symbols to detect the space-time redundancy. However, the approaches for identifying STBC-OFDM signals cannot be directly applied to SFBC-OFDM signals since the peaks of the cross-correlation functions between two adjacent OFDM symbols are difficult to detect. The previous works on the identification of SFBC-OFDM systems include \cite{blind_SFBC,My_paper_Globecom,My_paper_TWC2,My_paper_TVT}. In \cite{blind_SFBC}, the idea of detecting the peak of the cross-correlation function between two receive antennas is extended to identify SFBC-OFDM signals, which uses a specific time-lag during the same OFDM symbol. In \cite{My_paper_Globecom,My_paper_TWC2}, we utilize a cross-correlation function between two receive antennas at adjacent OFDM subcarriers to improve the identification performance by detecting both the space and frequency redundancy. In \cite{My_paper_TVT}, we use the random matrix theory to identify 5 types of SFBCs by detecting the space-frequency redundancy at adjacent OFDM subcarriers. However, most previous works can only discriminate between a few MIMO schemes since they only consider detecting the presence of the redundancy. Specifically, a widely-used TD code, namely the frequency switched transmit diversity (FSTD)\cite{sesia2009lte,IEEE802_16}, and several non-orthogonal STBCs/SFBCs can not be discriminated by the previous non-likelihood-based methods. Although the likelihood-based method \cite{Likelihood_Based} can identify more STBCs, it does not work in a frequency-selective fading environment.

In the existing literature, the enumeration of the number of transmit antennas and identification of MIMO schemes are handled as two independent problems. The enumeration problem of the number of transmit antennas is formulated as the enumeration of independent channel pathways between transmit antennas and receive antennas in general. Previous works on the identification of the number of transmit antennas mainly fall into two classes, namely second- \cite{AIC_MDL,eigenvector_nt_est,PET,WME} and higher-order statistics-based methods \cite{HOM_TD_Nt_est}. Basically, the second-order statistics-based methods analyze eigenvalues or eigenvectors of the covariance matrix of the received signals to determine the number of transmit antennas by distinguishing between the signal and noise subspaces. These methods also fall under two categories, information-theoretic criteria-based algorithms\cite{AIC_MDL} and hypothesis-testing-based algorithms \cite{eigenvector_nt_est,PET,WME}. Methods in the first category determine the number of transmit antennas by minimizing the Kullback-Leibler distance metric. Reference \cite{AIC_MDL} introduces two classical calculations of the Kullback-Leibler metric,  the Akaike information criterion (AIC) and the minimum description length (MDL), for the enumeration of the number of transmit antennas. Methods in the second category transform the problem into a detection problem, which compares an elaborately constructed statistic with a threshold. Furthermore, for the class of higher-order statistics-based methods, the sole existing algorithm \cite{HOM_TD_Nt_est} constructs a fourth-order decision statistic of the received signals with only one receive antenna by using the feature of time-varying block fading channels. To the best of our knowledge, no method exists in the literature for the joint blind identification of the number of transmit antennas and MIMO schemes. It is our main goal in this paper to fill this research gap.

Artificial neural networks (ANN) have been applied to signal identification problems, such as automatic modulation classification (AMC) \cite{AMC_ANN,MIMOAMC_ANN,AMC_CNN,AMC_LSTM}, since they are suitable for non-linear fitting and classification problems and do not impose any restrictions on the input variables, unlike other prediction techniques. Traditional ANNs require expert features, while modern deep learning neural networks can directly learn the statistical features from training data. References\cite{AMC_CNN,AMC_LSTM} use deep learning neural networks on the raw in-phase and quadrature phase (IQ) data to solve the AMC problem. However, their goal is to classify the single-antenna system. For the MIMO system, it is difficult to directly employ deep learning on the raw IQ data since MIMO overlapped signals destroy the statistical features. Given that the feed-forward neural network (FNN) is a popular family of ANN owing to its simple structure and strong fitting ability, it is used to develop high performance signal identification solutions \cite{AMC_ANN, MIMOAMC_ANN}.

As mentioned earlier, the quantification of the space-time/frequency redundancy can classify more STBCs/SFBCs since the redundancy of some STBCs/SFBCs is in the same location. On the other hand, this quantification is also needed in the enumeration of the number of transmit antennas. In this paper, a subspace-rank feature-based joint blind identification algorithm of the number of transmit antennas and MIMO schemes is proposed. Three different subspace-rank features for the number of transmit antennas and redundancy are derived from the eigenvalue analysis of the covariance matrix of the received signals at adjacent symbols or OFDM symbols/subcarriers with multiple receive antennas. A Gerschgorin radii-based method and an FNN are applied to calculate these features, and a minimal weighted norm-1 distance metric is proposed to determine the number of transmit antennas and MIMO schemes. The proposed algorithm does not require \textit{a priori} knowledge of the signal parameters, such as channel coefficients, modulation type or noise power.

The main contributions of this paper are the following:
\begin{itemize}
\item
The proposed algorithm jointly identifies the number of transmit antennas and MIMO schemes, which has not been considered in the previous works.
\item
The scenarios of single-carrier and OFDM, including STBC-OFDM and SFBC-OFDM are all investigated in this paper, unlike previous works.
\item
Unlike the existing algorithms, more STBC/SFBC types, such as the orthogonal STBCs/SFBCs (OSBC) with the same rate, FSTD, quasi-orthogonal STBC/SFBC (QOSBC) and non-orthogonal STBCs/SFBCs (SBC), are identified by the proposed algorithm thanks to the analysis of subspace-rank features.
\item
A Gerschgorin radii-based method and an FNN are efficiently combined to calculate the subspace-rank features. Furthermore, we extend the investigation to OFDM systems.
\item
The computational complexity of the proposed algorithm is analyzed and shown to be comparable to the enumeration algorithm of the number of transmit antennas in \cite{AIC_MDL} or identification algorithm of MIMO schemes in \cite{Likelihood_Based}.
\item
Simulation results are presented to demonstrate the viability of the proposed algorithm both in single-carrier and OFDM systems, with different system parameters.
\end{itemize}

This paper is organized as follows. In Section II, the system model is introduced. Then, Section III derives the three subspace-rank features. The proposed algorithm is described in Section IV. The simulations results are presented in Section V. Finally, conclusions are drawn in Section VI.

{\bf Notation:} The following notation is used throughout the paper. The superscripts ${   (\cdot )  ^ * }$, ${   (\cdot)  ^ {T} }$ and ${   (\cdot)   ^ {H} }$ denote the complex conjugate, transposition and conjugate transposition, respectively.  $\Pr \left( B \right)$ represents the probability of the event $B$. ${\rm{E}}\left[  \cdot  \right]$ indicates the statistical expectation. $\Re \left\{  \cdot  \right\}$ and $\Im \left\{  \cdot  \right\}$ denote the real and imaginary parts, respectively. $\bf{I}$, $\bf{0}$ and $\bf{O}$ denote the identity matrix, zero vector and zero matrix, respectively. $\mathbb{N}$, $\mathbb{Z}^+$ and $\mathbb{C}$ are the set of natural numbers, positive integers and complex numbers, respectively.  The notation ${\rm{card}}(A)$ denotes the cardinality of the set $A$. ${\rm Tr} (\cdot)$ denotes the trace of a matrix. Conventionally, $e$ and $\rm{log}$ denote the Euler constant and natural logarithm, respectively. Finally, ${\cal O}(\cdot)$ denotes the complexity order.

\section{System Model}

\subsection{Signal Model of MIMO Single-Carrier System}

\begin{figure*}
  \centering
  \includegraphics[width=0.98\textwidth,height=0.5\textheight]{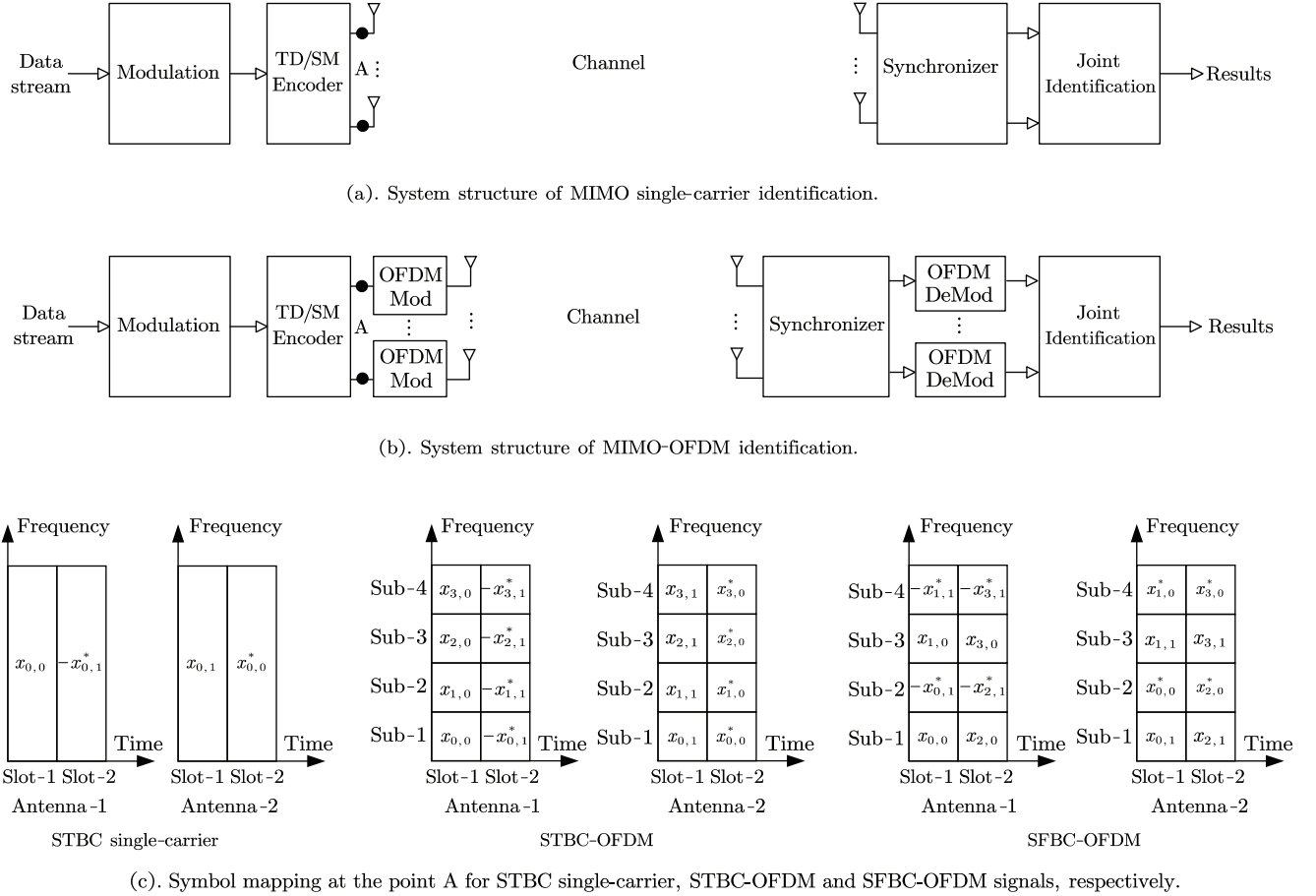}\\
  \caption{System structures and signal mappings of STBC/SFBC.}\label{fig1}
\end{figure*}

We consider a MIMO single-carrier wireless communication system employing TD or SM with ${N_t}$ transmit antennas and ${N_r}$ $( {{N_r} > {N_t}}  )$ receive antennas, as shown in Fig. \ref{fig1} (a). As a special case, single-antenna systems are also considered. The transmitted data symbols are drawn from an ${M}$-PSK (Phase-Shift-Keying) or ${M}$-QAM (Quadrature Amplitude Modulation), $M \ge 4$, signal constellation. Subsequently, the modulated symbol stream is parsed into a data block of ${N_s}$ symbols, denoted by the vector ${{\bf{x}}_b} = {[ {{x_{b,0}}, \cdots ,{x_{b,{N_s} - 1}}} ]^T}$ ($b \in \mathbb{N}$).  A TD/SM encoder takes the row of an $N_t \times T$ codeword matrix, denoted by ${\bf{C}}( {{{\bf{x}}_b}} )$, to span $T$ consecutive time slots and maps every column of the matrix into $N_t$ different transmit antennas. In this paper, the codewords include the single-antenna, Alamouti (AL), SM, 7 types of OSBC\cite{STBC_Tarokh,STBC_Ganesan,STBC_4_ant}, one type of QOSBC\cite{QOSTBC}, FSTD in LTE\cite{sesia2009lte} and 3 types of SBC in WiMAX\cite{IEEE802_16} (see Appendix A). Then, mapped signals are transmitted after the pulse shaping and carrier modulation operations. 

The receiver is assumed to successfully synchronize the received signals at the beginning to simplify the analysis; however, later we analyze the sensitivity to model mismatches in Section V. We construct an ${N_t} \times 1$ transmit vector and an ${N_r} \times 1$ receive vector, denoted by ${\bf{s}}(n)$ and ${\bf{y}}(n)$, which represent the transmitted and received signals at the $n$-th ($n \in \mathbb{Z}^+$) time slot, respectively. The channel is assumed to be flat-fading and characterized by an ${N_r} \times {N_t}$ matrix of Rayleigh fading coefficients, denoted by
\begin{equation}
{{\bf{H}}} =  \left[ {\begin{array}{*{20}{c}}
{h^{( {1,1} )}}& \cdots &{h^{( {{N_t},1} )}}\\
 \vdots & \ddots & \vdots \\
{h^{( {1,{N_r}} )}}& \cdots &{h^{( {{N_t},{N_r}} )}}
\end{array}}  \right] \label{eq1}
\end{equation}
where ${h^{( {f_1,f_2} )}}$ represents the channel coefficient between the $f_1$-th transmit antenna and $f_2$-th receive antenna. The channel matrix ${{\bf{H}}}$ is assumed to be of full-column rank and the channel gains remain constant over the observation interval. Then, the $n$-th received signal is described by the following model
\begin{equation}
{{\bf{y}}}( n ) = {{\bf{H}}}  {{\bf{s}}}( n ) + {{\bf{w}}}( n ) \label{eq2}
\end{equation}
where the vector ${\bf{w}} \left( n\right) $ represents a white Gaussian noise vector with zero-mean and covariance $\sigma _w^2{{\bf{I}}_{{N_r}}}$.
The first processed sample is assumed to be the start of a TD code block, which allows simplifications of the following mathematical expressions. However, extensions of the proposed methods can be easily obtained when this assumption does not hold.

\subsection{Signal Model of MIMO-OFDM System}
\subsubsection{STBC-OFDM System}

Consider a MIMO-OFDM wireless communication system with $N$ subcarriers and a cyclic prefix (CP) of length $\nu$, as shown in Fig. \ref{fig1} (b). Different from the single-carrier system, the TD/SM encoder puts $N$ data blocks, denoted by $ {{\bf{x}}_b , \cdots ,{\bf{x}}_{b + N - 1} } $, on $N$ consecutive subcarriers with the same operation as the single-carrier system. At the receiver side, assume that the carrier type is successfully estimated which can be achieved based on the cyclic cumulant \cite{OFDM_v_single}. the received OFDM symbol is converted into a frequency-domain block via an $N$-point fast Fourier transform (FFT) after removing the CP. We construct an ${N_t} \times 1$ transmit vector and an ${N_r} \times 1$ receive vector, denoted by ${\bf{s}}_k(n)$, and ${\bf{y}}_k(n)$, which represent the transmitted and received signals at the $n$-th time slot and $k$-th ($1 \leq k \leq N$) subcarrier, respectively.
The channel is assumed to be frequency-selective fading and the $k$-th subchannel is characterized by an ${N_r} \times {N_t}$ full-rank matrix of fading coefficients, denoted by ${\bf{H}}_k$. Then, the $n$-th received signal at the $k$-th subcarrier is described by the following model
\begin{equation}
{{\bf{y}}_k}( n ) = {{\bf{H}}_k}  {{\bf{s}}_k}( n ) + {{\bf{w}}_k}( n ) \label{eq4}
\end{equation}
where the ${N_r} \times 1$ vector ${\bf{w}}_k \left( n\right)$ represents a frequency-domain white Gaussian noise vector at the $k$-th subcarrier.

\subsubsection{SFBC-OFDM System}

The SFBC-OFDM system model is similar to the STBC-OFDM system model, with the difference that the SFBC encoder takes the row of the codeword matrix ${\bf{C}}( {{{\bf{x}}_b}} )$ to span $T$ consecutive subcarriers directly ($T$ is the number of the column of ${\bf{C}}( {{{\bf{x}}_b}} )$). The three mappings are shown in Fig. \ref{fig1} (c).

\section{Subspace-Rank Features of Different Numbers of Transmit Antennas and MIMO Schemes}
In this section, three subspace-rank features are defined as discriminating features of different MIMO schemes, by considering the dimension of the subspace of restructured received signals. Without loss of generality, we consider the single-carrier system first and then extend the analysis to the OFDM system.

\subsection{Number of Transmit Antennas Feature}
Let us construct a time-domain receive window to observe the received signals at adjacent time slots. The window length is set to two since it is the finest granularity to quantify the features of different MIMO schemes, while a larger window length results in failure to distinguish many MIMO schemes. By using \eqref{eq2}, the $l$-th received signal block inside the window is expressed as
\begin{equation}
{{\bf{Y}}} ( l  ) = {{\bf{H}}}  {{\bf{S}}} ( l  ) + {{\bf{W}}} ( l  ) \label{eq5}
\end{equation}
where the transmitted signal block of adjacent time slots is denoted by an ${N_t} \times 2$ matrix ${\bf{S}}( l ) = [ {{\bf{s}}( l ),{\bf{s}}( {l + 1} )} ]$, and the noise vector is denoted by ${{\bf{W}}} ( l ) = [ {{{\bf{w}}} ( l ),{{\bf{w}}} ( l+1 )}  ]$. The covariance matrix of the received block is
\begin{equation}
{{\bf{\Sigma }}_{\bf{Y}}} ( l  ) = {\rm{E}} [ {{{\bf{Y}}} ( l  ){\bf{Y}}^H ( l  )}  ] = {{\bf{H}}}  {{\bf{\Sigma }}_{\bf{S}}} ( l  )  {\bf{H}}^H + 2\sigma _w^2{{\bf{I}}_{{N_r}}}   \label{eq6}
\end{equation}
where ${{\bf{\Sigma }}_{\bf{S}}} ( l  ) = {\rm{E}} [ {{{\bf{S}}} ( l  ){\bf{S}}^H ( l  )}  ]$ is the covariance matrix of the $l$-th transmitted block. Denoting the eigenvalues of ${{\bf{\Sigma }}_{\bf{Y}}} ( l )$ by in descending order, we have the eigenvalues $\varphi _1 ( l  ) \ge \varphi _2 ( l  ) \cdots  \ge \varphi _{{N_r}} ( l  )$.

\emph{Proposition 1:} The smallest ${N_r} - {N_t}$ ordered eigenvalues of ${{\bf{\Sigma }}_{\bf{Y}}}$ are all equal to $2 \sigma _w^2$, i.e., $\varphi _{{N_t} + 1} =  \cdots  = \varphi _{{N_r}} = 2 \sigma _w^2$ if the rank of ${{\bf{\Sigma }}_{\bf{S}}}$ is $N_t$.

\emph{Proof:} See Appendix B.

The corresponding eigenvectors of the eigenvalues $ \varphi _1\left( l \right) ,\varphi _2\left( l \right) \cdots ,\varphi _{N_r-N_t}\left( l \right)  $ form a basis for the signal subspace. Define the following subset of signal subspace eigenvalues of ${{\bf{\Sigma }}_{\bf{Y}}} ( l  )$
\begin{equation}
{{\cal A}_l} =  \left\{ \varphi _1\left( l \right) ,\varphi _2\left( l \right) \cdots ,\varphi _{N_r-N_t}\left( l \right) \right\}. \label{eq8}
\end{equation}
By sliding the window, the cardinality of the set ${\cal A}_l$ with even subscript ($l = 2m$, $m \in \mathbb{Z}^+$) can be used as the first type of the subspace-rank feature since the cardinality of ${\cal A}_{2m-1}$ for FSTD is equal to two. Therefore, we define this cardinality as the number of transmit-antenna feature (NTAF), denoted by
\begin{equation}
{\alpha } =  {\rm card} ({{\cal A}_l}), \quad l = 2m, m \in \mathbb{Z}^+. \label{eq9}
\end{equation}
Note that the NTAF, $\alpha$, is the discriminating feature for different numbers of transmit antennas.

\subsection{Number of Independent Complex Symbols Feature}
Let us vectorize the $l$-th signal and noise block inside the window in \eqref{eq5} as follows
\begin{gather}
\mathbf{\bar{y}}\left( l \right) =\left[ \begin{array}{c}
	\mathbf{y}\left( l \right)\\
	\mathbf{y}\left( l+1 \right)\\
\end{array} \right], \quad \mathbf{\bar{s}}\left( l \right) =\left[ \begin{array}{c}
	\mathbf{s}\left( l \right)\\
	\mathbf{s}\left( l+1 \right)\\
\end{array} \right] \notag \\
 \mathbf{\bar{w}}\left( l \right) =\left[ \begin{array}{c}
	\mathbf{w}\left( l \right)\\
	\mathbf{w}\left( l+1 \right)\\
\end{array} \right] . \label{eq10}
\end{gather}
Then, the $l$-th received vectorized block is rewritten as
\begin{equation}
{{{\bf{\bar y}}}} ( l  ) = {{{\bf{\bar H}}}}  {{{\bf{\bar s}}}} ( l  ) + {{{\bf{\bar w}}}} ( l  ) \label{eq12}
\end{equation}
where the $2{N_r} \times 2{N_t}$ matrix ${{{\bf{\bar H}}}}$ is
\begin{equation}
{{{\bf{\bar H}}}} = \left[ {\begin{array}{*{20}{c}}
{{{\bf{H}}}}&{\bf{O}}\\
{\bf{O}}&{{{\bf{H}}}}
\end{array}}  \right]. \label{eq13}
\end{equation}
Like \eqref{eq6}, the covariance matrix of the $l$-th received vectorized block is
\begin{equation}
{{\bf{\Sigma }}_{{\bf{\bar y}}}} ( l  ) = {\rm{E}} [ {{{{\bf{\bar y}}}} ( l  ){\bf{\bar y}}^H ( l  )} ] = {{{\bf{\bar H}}}}  {{\bf{\Sigma }}_{{{{\bf{\bar s}}}}}} ( l  )  {\bf{\bar H}}^H + \sigma _w^2{{\bf{I}}_{2{N_r}}} \label{eq14}
\end{equation}
where ${{\bf{\Sigma }}_{{{{\bf{\bar s}}}}}} ( l  ) = {\rm{E}} [ {{{{\bf{\bar s}}}} ( l  ){\bf{\bar s}}^H ( l )}  ]$ is the covariance matrix of the $l$-th transmitted vectorized block. The eigenvalues of ${{\bf{\Sigma }}_{{\bf{\bar y}}}} ( l  )$ are $\phi  _1 ( l  ) \ge \phi _2 ( l  ) \cdots  \ge \phi _{2{N_r}} ( l  )$.

Next, we define the notion of linearly independent random symbols. Let ${\bf{v}}_1, \cdots, {\bf{v}}_n$ be the vector observations of the random variables $v_1, \cdots, v_n$. Then, we define $v_1, \cdots, v_n$ as linearly independent random symbols if the equation $c_1{\bf{v}}_1 + c_2{\bf{v}}_2 + \cdots + c_n{\bf{v}}_n = 0$ can only be satisfied by $c_i = 0$ for $i = 1, \cdots, n$ \cite{linear_indep_v}. For example, assuming ${\bf{u}} =  [ {v_1},{v_2},{{v_3} /{\sqrt 2 }}, - v_2^ * ,v_1^ * ,{{v_3} /{\sqrt 2 }}  ]^T$ whose elements are complex random variables, then ${\bf{u}}$ has five linearly independent complex random symbols, i.e., ${v_1}$, ${v_2}$, ${{v_3} /{\sqrt 2 }}$, ${v_1^ *}$, ${- v_2^ * }$. Assuming $\mathbf{u}= [ 
	\Re \left( v_1 \right),	\Im \left( v_1 \right), 	\Re \left( v_2 \right),		\Im \left( v_2 \right),		\Re \left( -v_{2}^{*} \right),		\Im \left( -v_{2}^{*} \right),  \Re \left( v_{1}^{*} \right), \\ \Im \left( v_{1}^{*} \right) ] ^T$,
then $\mathbf{u}$ has four linearly independent real random symbols, i.e., ${\Re  ( {{v_1}}  )}$, ${\Im  ( {{v_1}}  )}$, ${\Re  ( {{v_2}}  )}$, ${\Im  ( {{v_2}}  )}$.

\emph{Proposition 2:} Assume that $M_c$ is the number of linearly independent complex random symbols of a transmitted vectorized block. The smallest ${2N_r} - {M_c}$ ordered eigenvalues of ${{\bf{\Sigma }}_{{\bf{\bar y}}}}$ are all equal to $ \sigma _w^2$, i.e., $\phi _{{M_c} + 1} =  \cdots  = \phi _{2{N_r}} = \sigma _w^2$.

\emph{Proof:} See Appendix B.

Similarly, let us define a subset of the signal subspace eigenvalues of ${{\bf{\Sigma }}_{{\bf{\bar y}}}}$ as
\begin{equation}
{\cal B}_l =  \left\{ \phi _1\left( l \right) ,\phi _2\left( l \right) \cdots ,\phi _{M_c}\left( l \right) \right\} . \label{eq17}
\end{equation}
By sliding the window, we can select the cardinalities of three sets, namely ${\cal B}_{l_1}$, ${\cal B}_{l_2}$ and ${\cal B}_{l_3}$ with subscripts $l_1 = 4m-3$, $l_2=4m-2$, and $l_3=4m-1$ ($m \in \mathbb{Z}^+$), respectively, as the second type of subspace-rank features to identify more MIMO schemes in the MIMO scheme pool. Then, we define three linearly independent complex random-symbol features (ICSF), which represent the cardinalities of the three sets, denoted by
\begin{subequations} \label{eq18}
\begin{align}
&\beta _1=\text{card}\left( \mathcal{B}_{l_1} \right) ,\quad l_1=4m-3,m\in \mathbb{Z}^+ \label{eq18a}\\
&\beta _2=\text{card}\left( \mathcal{B}_{l_2} \right) ,\quad l_2=4m-2,m\in \mathbb{Z}^+ \label{eq18b}\\
&\beta _3=\text{card}\left( \mathcal{B}_{l_3} \right) ,\quad l_3=4m-1,m\in \mathbb{Z}^+.\label{eq18c}
\end{align}
\end{subequations}

\subsection{Number of Independent Real Symbols Feature}

By stacking the real and imaginary parts of the block inside the window in \eqref{eq10}, we obtain a transmitted, a received stacked block and the noise block as
\begin{gather}
\mathbf{\tilde{y}}\left( l \right) =\left[ \begin{array}{c}
	\Re \left( \mathbf{y}\left( l \right) \right)\\
	\Im \left( \mathbf{y}\left( l \right) \right)\\
	\Re \left( \mathbf{y}\left( l+1 \right) \right)\\
	\Im \left( \mathbf{y}\left( l+1 \right) \right)\\
\end{array} \right], \,\,\mathbf{\tilde{s}}\left( l \right) =\left[ \begin{array}{c}
	\Re \left( \mathbf{s}\left( l \right) \right)\\
	\Im \left( \mathbf{s}\left( l \right) \right)\\
	\Re \left( \mathbf{s}\left( l+1 \right) \right)\\
	\Im \left( \mathbf{s}\left( l+1 \right) \right)\\
\end{array} \right] \notag \\
\mathbf{\tilde{w}}\left( l \right) =\left[ \begin{array}{c}
	\Re \left( \mathbf{w}\left( l \right) \right)\\
	\Im \left( \mathbf{w}\left( l \right) \right)\\
	\Re \left( \mathbf{w}\left( l+1 \right) \right)\\
	\Im \left( \mathbf{w}\left( l+1 \right) \right)\\
\end{array} \right]. \label{eq19}
\end{gather}
Similarly, the $l$-th received stacked block is rewritten as
\begin{equation}
{{{\bf{\tilde y}}}} ( l  ) =  ( {{{\bf{I}}_2} \otimes {{{\bf{\tilde H}}}}}  ){{{\bf{\tilde s}}}} ( l  ) + {{{\bf{\tilde w}}}} ( l  ) \label{eq21}
\end{equation}
where the $2{N_r} \times 2{N_t}$ matrix ${{\bf{\tilde H}}}$ is given by
\begin{equation}
{{{\bf{\tilde H}}}} =  \left[ {\begin{array}{*{20}{c}}
{\Re  ( {{{\bf{H}}}}  )}&{ - \Im  ( {{{\bf{H}}}}  )}\\
{\Im  ( {{{\bf{H}}}}  )}&{\Re  ( {{{\bf{H}}}}  )}
\end{array}} \right ] \label{eq22}
\end{equation}
and $ \otimes $ corresponds to the Kronecker product. Then, the covariance matrix of ${\bf{\tilde y}} ( l  )$ is
\begin{align}
{{\bf{\Sigma }}_{{\bf{\tilde y}}}} ( l  ) &= {\rm{E}} [ {{{{\bf{\tilde y}}}} ( l  ){\bf{\tilde y}}^T ( l )}  ] \notag \\
 &=  ( {{{\bf{I}}_2} \otimes {{{\bf{\tilde H}}}}}  ){{\bf{\Sigma }}_{{\bf{\tilde s}}}} ( l  ) ( {{{\bf{I}}_2} \otimes {\bf{\tilde H}}^T}  ) + \frac{{\sigma _w^2}}{2}{{\bf{I}}_{4{N_r}}} \label{eq23}
\end{align}
where ${{\bf{\Sigma }}_{{\bf{\tilde s}}}} ( l  ) = {\rm{E}} [ {{{{\bf{\tilde s}}}} ( l  ){\bf{\tilde s}}^T ( l )}  ]$ is the covariance matrix of the $l$-th transmitted stacked block. The eigenvalues of ${{\bf{\Sigma }}_{{\bf{\tilde y}}}} ( l  )$ are $\eta _1 ( l  ) \ge \eta _2 ( l  ) \cdots  \ge \eta _{4{N_r}} ( l  )$.

\emph{Proposition 3:} Assume that $M_r$ is the number of linearly independent real random symbols of a transmitted stacked block. The smallest ${4N_r} - {M_r}$ ordered eigenvalues of ${{\bf{\Sigma }}_{{\bf{\tilde y}}}} ( l  )$ are all equal to $ {{\sigma _w^2}/ 2}$, i.e., $\eta _{{M_r} + 1} =  \cdots  = \eta _{4{N_r}}  = {{\sigma _w^2} /2}$.

\emph{Proof:}  See Appendix B.

Let us define the following subset of the signal subspace eigenvalues of ${{\bf{\Sigma }}_{{\bf{\tilde s}}}} ( l  )$ as
\begin{equation}
{\cal C}_l =  \left\{ \eta _1\left( l \right) ,\eta _2\left( l \right), \cdots ,\eta _{M_r}\left( l \right) \right\} . \label{eq27}
\end{equation}
By sliding the window, the cardinalities of the sets ${\cal C}_{l_1}$ and ${\cal C}_{l_2}$  with subscripts $l_1 = 4m-3$ and $l_2 = 4m-1$ ($m \in \mathbb{Z}^+$), respectively, are chosen to be the third type of subspace-rank features since the cardinalities of the other sets can not be used to identify several TD codes with different transmission rates. Then, we define two linearly independent real random-symbol feature (IRSF) representing these cardinalities given by
\begin{subequations} \label{eq28}
\begin{align}
&\gamma_1  =\text{card}\left( \mathcal{C}_{l_1} \right) ,\quad l_1=4m-3,m\in \mathbb{Z}^+ \label{eq28a}\\
&\gamma_2  =\text{card}\left( \mathcal{C}_{l_2} \right) ,\quad l_2=4m-1,m\in \mathbb{Z}^+. \label{eq28b}
\end{align}
\end{subequations} 

Actually, the ICSF and IRSF can quantify the space-time redundancy. For different MIMO schemes with the same number of transmit antennas, the transmitted signal block inside the window has more space-time redundancies, or in other words, the block transmits some identical symbols due to the structure of the codeword matrix. Therefore, smaller ICSFs and IRSFs values are calculated at the receiver since these symbols are linearly dependent.

All signal features using different numbers of transmit antennas and common MIMO schemes are listed, and a representative example are described in Appendix C.

\subsection{Subspace-Rank Features in the OFDM System}

\subsubsection{STBC-OFDM Case}
Since STBC-OFDM is a direct extension of STBC single-carrier, we can use the $l$-th received signal block at the $k$-th subcarrier, denoted by $\mathbf{Y}_l\left( n \right)$, to derive the NTAF, ICSFs and IRSFs as we described previously.


\subsubsection{SFBC-OFDM Case}
Regarding the SFBC-OFDM system, we construct and slide a frequency-domain receive window to observe the received signals at adjacent subcarriers and calculate the subspace-rank features.
Thus, the $l$-th received block, vectorized block and stacked block in \eqref{eq5}, \eqref{eq10} and \eqref{eq19} inside the window are, respectively, modified to
\begin{gather} 
 \mathbf{Y}_l\left( n \right) =\left[ \mathbf{y}_l\left( n \right) ,\mathbf{y}_{l+1}\left( n \right) \right] , \quad \mathbf{\bar{y}}_l\left( n \right) =\left[ \begin{array}{c}
	\mathbf{y}_l\left( n \right)\\
	\mathbf{y}_{l+1}\left( n \right)\\
\end{array} \right] \notag \\
\mathbf{\tilde{y}}_l\left( n \right) =\left[ \begin{array}{c}
	\Re \left( \mathbf{y}_l\left( n \right) \right)\\
	\vdots\\
	\Im \left( \mathbf{y}_{l+1}\left( n \right) \right)\\
\end{array} \right]. \label{eq32}
\end{gather}

\section{Proposed Blind Identification Algorithm}
In this section, we use a Gerschgorin radii-based method and an FNN to calculate the subspace-rank features and employ a minimal weighted norm-1 distance metric to discriminate between these features. Different from the original Gerschgorin radii-based method in \cite{Gerschgorin}, the radii of the circles are compressed after a similarity transformation, and then, an FNN is used to calculate the subspace-rank features. Additionally, extensions to the STBC-OFDM and SFBC-OFDM systems are proposed by combining the calculations from different subcarriers after an analysis of the channel response. 

\subsection{Proposed Algorithm for Single-Carrier System}

The covariance matrices estimators are given by
\begin{subequations} \label{eq34}
\begin{align}
&\mathbf{\hat{\Sigma}}_{\alpha }=\frac{1}{L/2-1}\sum_{m=1}^{L/2-1}{\mathbf{Y}\left( 2m \right) \cdot \mathbf{Y}^H\left( 2m \right)} \label{eq34a}\\
&\mathbf{\hat{\Sigma}}_{{\beta }_1}=\frac{1}{L/4}\sum_{m=1}^{L/4}{\mathbf{\bar{y}}\left( 4m-3 \right) \cdot \mathbf{\bar{y}}^H\left( 4m-3 \right)} \label{eq34b}\\
&\mathbf{\hat{\Sigma}}_{{\beta }_2}=\frac{1}{L/4}\sum_{m=1}^{L/4}{\mathbf{\bar{y}}\left( 4m-2 \right) \cdot \mathbf{\bar{y}}^H\left( 4m-2 \right)}\label{eq34c}\\
&\mathbf{\hat{\Sigma}}_{{\beta }_3}=\frac{1}{L/4}\sum_{m=1}^{L/4}{\mathbf{\bar{y}}\left( 4m-1 \right) \cdot \mathbf{\bar{y}}^H\left( 4m-1 \right)} \label{eq34d}\\
&\mathbf{\hat{\Sigma}}_{{\gamma_1 }}=\frac{1}{L/4}\sum_{m=1}^{L/4}{\mathbf{\tilde{y}}\left( 4m-3 \right) \cdot \mathbf{\tilde{y}}^T\left( 4m-3 \right)} \label{eq34e}\\
&\mathbf{\hat{\Sigma}}_{{\gamma_2 }}=\frac{1}{L/4}\sum_{m=1}^{L/4}{\mathbf{\tilde{y}}\left( 4m-1 \right) \cdot \mathbf{\tilde{y}}^T\left( 4m-1 \right)} \label{eq34f}
\end{align}
\end{subequations}
where $L$ is the number of symbols. For convenience, we employ a common notation, ${\bf{\hat{\Sigma }}}$, to represent the estimated covariance matrices in \eqref{eq34}.
Assume that ${\bf{\hat{\Sigma }}}$ is a $J \times J$ matrix. First, we partition the estimated covariance matrix as
\begin{equation}
{\bf{\hat \Sigma }} = \left [ {\begin{array}{*{20}{c}}
{{a_{11}}}&{{a_{12}}}& \cdots &{{a_{1J}}}\\
{{a_{21}}}&{{a_{22}}}& \cdots &{{a_{2J}}}\\
 \vdots & \vdots & \ddots & \vdots \\
{{a_{J1}}}&{{a_{J2}}}& \cdots &{{a_{JJ}}}
\end{array}}  \right] = \left [ {\begin{array}{*{20}{c}}
{{{\bf{\Sigma }}_1}}&{\bf{a}}\\
{{{\bf{a}}^H}}&{{a_{JJ}}}
\end{array}} \right] \label{eq35}
\end{equation}
where the reduced covariance matrix ${{{\bf{\Sigma }}_1}}$ is a $(J - 1) \times (J - 1)$ leading principal submatrix of ${{\bf{\hat \Sigma }}}$ obtained by removing the last row and column of ${{\bf{\hat \Sigma }}}$. Then, the reduced covariance matrix ${{{\bf{\Sigma }}_1}}$ can be decomposed by its eigenstructure as follows
\begin{equation}
{{\bf{\Sigma }}_1} = {{\bf{Q}}_1}{{\bf{\Lambda }}_1}{\bf{Q}}_1^H \label{eq36}
\end{equation}
where ${{\bf{\Lambda }}_1}$ is a diagonal matrix constructed from the eigenvalues of ${{{\bf{\Sigma }}_1}}$ as
\begin{equation}
{{\bf{\Lambda }}_1} = {\rm{diag}} ( {{\mu _1},{\mu _2}, \cdots ,{\mu _{J - 1}}}  ) \label{eq37}
\end{equation}
 and ${{\bf{Q}}_1}$ is a $(J - 1) \times (J - 1)$ unitary matrix formed by the corresponding eigenvectors as follows
\begin{equation}
{{\bf{Q}}_1} =  \left[ {{{\bf{q}}_1},{{\bf{q}}_2}, \cdots ,{{\bf{q}}_{J - 1}}}  \right]. \label{eq38}
\end{equation}
Then, we construct the following $J \times J$ unitary transformation matrix  
\begin{equation}
{{\bf{Q}}_2} =  \left[ {\begin{array}{*{20}{c}}
{{{\bf{Q}}_1}}&{\bf{0}}\\
{{{\bf{0}}^T}}&1
\end{array}}  \right] \label{eq39}
\end{equation}
where ${{\bf{Q}}_2}{\bf{Q}}_2^H = {\bf{I}}$. The transformation of ${{\bf{\hat \Sigma }}}$ is
\begin{align}
\mathbf{R}=\mathbf{Q}_2\mathbf{\hat{\Sigma}Q}_{2}^{H} &=\left[ \begin{matrix}
	\mathbf{\Lambda }_1&		\mathbf{Q}_{1}^{H}\mathbf{a}\\
	\mathbf{a}^H\mathbf{Q}_1&		a_{JJ}\\
\end{matrix} \right]  \notag \\
&=\left[ \begin{matrix}
	\mu _1&		0&		\cdots&		0&		\rho _1\\
	0&		\mu _2&		\cdots&		0&		\rho _2\\
	\vdots&		\vdots&		\ddots&		\vdots&		\vdots\\
	0&		0&		\cdots&		\mu _{J-1}&		\rho _{J-1}\\
	\rho _{1}^{*}&		\rho _{2}^{*}&		\cdots&		\rho _{J-1}^{*}&		a_{JJ}\\
\end{matrix} \right]       \label{eq40}
\end{align}
where ${\rho _i} = {\bf{q}}_i^H{\bf{a}}$ for $i = 1, 2, \cdots , J - 1$. Thus, the radius of the $i$-th Gerschgorin circle is
\begin{equation}
{r_i} =  | {{\rho _i}}  | =  | {{\bf{q}}_i^H{\bf{a}}}| .  \label{eq41}
\end{equation}
In order to scale the radii of Gerschgorin circles in proportion, we construct the diagonal matrix
\begin{equation}
{\bf{P}} = {\rm{diag}} ( {{\mu _1},{\mu _2}, \cdots ,{\mu _{J - 1}},{\mu _{J}}}  ) \label{eq42}
\end{equation}
where $\mu _{J}$ is the mean of the eigenvalues $\mu_1, \cdots,  \mu_{J-1}$ given by
\begin{equation}
{\mu _J} = \frac{1}{{J - 1}}\sum\limits_{i = 1}^{J - 1} {{\mu _i}} . \label{eq43}
\end{equation}
The matrix ${\bf{R}}$ can be similarly transformed into
\begin{align}
\mathbf{R'}&=\mathbf{PRP}^{-1} \notag \\
&=\left[ \begin{matrix}
	\mu _1&		0&		\cdots&		0&		\frac{\mu _1}{\mu _J}\rho _1\\
	0&		\mu _2&		\cdots&		0&		\frac{\mu _2}{\mu _J}\rho _2\\
	\vdots&		\vdots&		\ddots&		\vdots&		\vdots\\
	0&		0&		\cdots&		\mu _{J-1}&		\frac{\mu _{J-1}}{\mu _J}\rho _{J-1}\\
	\frac{\mu _J}{\mu _1}\rho _{1}^{*}&		\frac{\mu _L}{\mu _2}\rho _{2}^{*}&		\cdots&		\frac{\mu _J}{\mu _{J-1}}\rho _{J-1}^{*}&		a_{JJ}\\
\end{matrix} \right] .    \label{eq44}
\end{align}
Practically, the centers of the Gerschgorin circles are fixed while their radii are relatively squeezed by the factor ${\mu_i}/{\mu_J}$.  The Gerschgorin circles of the noise subspace are squeezed more than those of the signals since the noise circles radii are generally smaller than $\mu _J$. Then, the radii of the compressed Gerschgorin circles for ${i = 1, \cdots ,J - 1 }$, are denoted by
\begin{equation}
{R_i} = \left| {\frac{{{\mu _i}{\rho _i}}}{{{\mu _J}}}} \right| = \frac{{{\mu _i}}}{{{\mu _J}}}{r_i}. \label{eq45}
\end{equation}

\begin{figure}
\centering
\includegraphics[width=.45\textwidth]{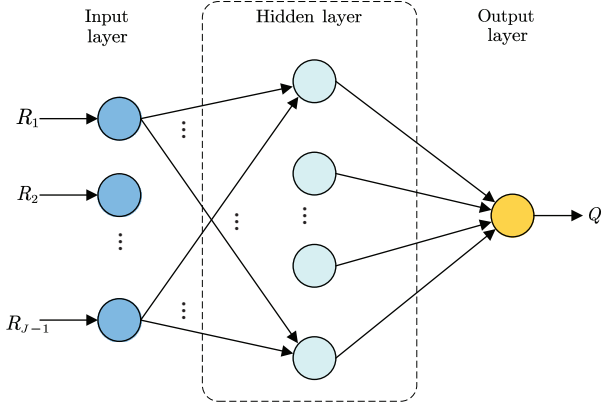}
\caption{FNN for the calculation of the Gerschgorin radii.}
\label{NN}
\end{figure}

After extracting the radii of the compressed Gerschgorin circles, the identification problem can be considered as a fitting problem of the Gerschgorin circles which determines how many Gerschgorin circles are those of the signal subspace. However, the radii of the Gerschgorin circles of the signal and noise subspaces have a wide range under different conditions, including signal-to-noise ratio (SNR), the number of processed samples and number of receive antennas, which results in the non-linearity between the inputs and outputs of the learning system. 

In this paper, since an FNN can fit any finite input-output mapping problem and has a simple structure, we use a three-layer FNN, as shown in Fig. \ref{NN}, to determine the number of the signal-subspace Gerschgorin circles. The FNN includes an input layer, a hidden layer and an output layer.   The hidden layer has 10-20 neurons using the sigmoid transfer function while the output layer only has one linear neuron. After the SNR is normalized, the Levenberg-Marquardt backpropagation algorithm \cite{more1978levenberg} is utilized to train the FNN by using test data. To avoid overfitting, we use a large set of data to train the FNNs, which is described later on in Section V.
Then, the feature value is a fitting function of the radii of compressed Gerschgorin circles given by
\begin{equation}
Q=f\left( R_1,R_2,\cdots ,R_{J-1} \right) . \label{eq46}
\end{equation}
The quantity $Q$ represents the calculated feature, $\hat{\alpha}$, $\hat{\beta}_1$, $\hat{\beta}_2$, $\hat{\beta}_3$, $\hat{\gamma_1}$ or $\hat{\gamma_2}$ depending on the corresponding covariance matrix in Equations \eqref{eq34a}-\eqref{eq34f}. Since the sizes of the covariance matrices and the eigenvalues after decomposition have large differences for the three subspace-rank features, the numbers and values (the distributions of values of the radii) of the FNN inputs are significantly different for different features. To enhance performance, we use three trained FNNs to determine $\{ \hat{\alpha} \}$, $\{ \hat{\beta}_1, \hat{\beta}_2, \hat{\beta}_3 \}$ and $\{ \hat{\gamma_1}, \hat{\gamma_2} \}$, respectively.
Finally, since the MIMO scheme $C$ contains the information on the number of transmit antennas $N_t$, they are simultaneously determined by a minimal weighted norm-1 distance metric given by
\begin{align}
\hat{N}_t,\hat{C}=\text{arg}\underset{C\in \left\{ \text{CODE} \right\}}{\min} (& 24 \cdot \left| \hat{\alpha}-\alpha \right|+4 \cdot \sum_i{\left| \hat{\beta}_i-\beta _i \right|} \notag \\
&+3 \cdot \sum_i{\left| \hat{\gamma}_i-\gamma _i \right|} )  \label{eq47}
\end{align}
where the notation \{CODE\} refers to the set of all schemes listed in Table \ref{table1} (see Appendix C). The reason for employing a norm-1 distance metric is that it is more robust against outliers than other distance metrics \cite{L1_distance}. The weighting coefficients are chosen to eliminate the bias caused by the features with larger values and equal to the least common multiple of the NTAF, sum of ICSFs and sum of IRSFs for single-antenna, which is equal to 24, divided by these values resulting in the weighting coefficients of 1, 6, 8, respectively.\footnote{ It is unfair for the coefficient of the NTAF if the MIMO scheme with the space-time redundancy is chosen here, since the NTAF does not quantify the space-time redundancy and has the same value for different MIMO schemes with the same number of transmit antennas.} For clarity, the main steps of the proposed algorithm are summarized subsequently.
 \begin{algorithm}
 \renewcommand{\thealgorithm}{}
 \caption{}
 \begin{algorithmic}[1]
 \renewcommand{\algorithmicrequire}{\textbf{Input: }}
 \renewcommand{\algorithmicensure}{\textbf{Output:}}
 \REQUIRE The observed sequence $\bf{y}$.
 \ENSURE  The number of transmit antennas $\hat{N}_t$ and MIMO scheme $\hat{C}$.
 \STATE{ Construct the received block ${\bf{Y}}$, the vectorized block ${\bf{\bar y}}$ and the stacked block ${\bf{\tilde y}}$ using \eqref{eq5}, \eqref{eq10} and \eqref{eq19}, respectively.}
 \STATE{ Compute the covariance matrices defined in \eqref{eq34}.} 
 \STATE{ Compute ${\bf{\Lambda }}_1$ and ${\bf{Q}}_1$ using the eigenvalue decomposition. } 
 \STATE{ Compute the radii of the original Gerschgorin circles ${r_i}$ using \eqref{eq41}.}
 \STATE{ Compute the radii of the compressed Gerschgorin circles ${R_i}$ using \eqref{eq45}.}
 \STATE{ Calculate the subspace-rank features $\{ \hat{\alpha} \}$, $\{ \hat{\beta}_1, \hat{\beta}_2, \hat{\beta}_3 \}$ and $\{ \hat{\gamma_1}, \hat{\gamma_2} \}$ by three trained FNNs, respectively.}
 \STATE{ Compute $\hat{N}_t$ and $\hat{C}$ using the weighted norm-1 distance formula in \eqref{eq47}.}
\RETURN { $\hat{N}_t$, $\hat{C}$.}
 \end{algorithmic}
 \end{algorithm}
 
%

\subsection{Extension to OFDM Systems}
For an OFDM system, each frequency subchannel can be reasonably assumed to be a quasi-static flat-fading channel since the subchannel width is designed to be much less than the channel's coherence bandwidth constraint. The frequency responses of adjacent subchannels are close to each other, especially when increasing the number of subchannels under a given total bandwidth constraint. Therefore, we can rewrite the subchannel frequency response at the $(k+1)$-th subcarrier as
\begin{equation}
{{\bf{H}}_{k + 1}} = {{\bf{H}}_k} + \Delta {\bf{H}} \label{eq48}
\end{equation}
where the tiny increment, $\Delta {\bf{H}}$, has the form
\begin{equation}
\Delta {\bf{H}} =  \left[ {\begin{array}{*{20}{c}}
{\Delta {h^{ ( {1,1}  )}}}& \cdots &{\Delta {h^{ ( {{N_t},1}  )}}}\\
 \vdots & \ddots & \vdots \\
{\Delta {h^{ ( {1,{N_r}}  )}}}& \cdots &{\Delta {h^{ ( {{N_t},{N_r}}  )}}}
\end{array}}  \right]. \label{eq49}
\end{equation}

\begin{figure}
\centering
\includegraphics[width=.5\textwidth]{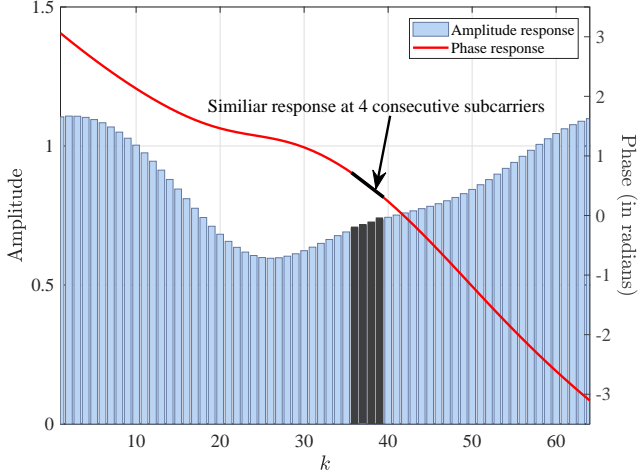}
\caption{Channel response of a frequency-selective fading channel consisting of 3 independent taps for $N=64$. }
\label{fig2}
\end{figure}

Fig. \ref{fig2} shows the frequency response for a frequency-selective fading channel which consists of 3 independent taps with an exponential power delay profile\cite{Blind_MIMO_OFDM_SM_AL}, $\sigma _t^2 = {e^{ - {t / 5}}}$. In this figure, the number of subchannels is set to 64, and the two y-axes represent  the amplitude and phase responses, respectively. From the figure, we can reasonably assume that four consecutive subcarriers have similar responses, expressed as
\begin{equation}
\mathbf{H}_k\approx \mathbf{H}_{k+1}\approx \mathbf{H}_{k+2}\approx \mathbf{H}_{k+3}. \label{eq50}
\end{equation}

\subsubsection{STBC-OFDM Case}
For convenience, let us use a new variable to rearrange the subscript indices of subcarriers, denoted by $p=\lceil {k}/{4} \rceil$, where $\lceil \cdot \rceil$ represents the ceiling function. Assume that $N_b$ OFDM symbols are observed at the receiver. Signals at four consecutive subcarriers are serially incorporated into a data block and the $p$-th data block is denoted by
\begin{equation}
\mathbf{\dot{y}}_p=\left[ \mathbf{y}_k\left( 1 \right) ,\cdots ,\mathbf{y}_k\left( N_b \right) ,\mathbf{y}_{k+1}\left( 1 \right), \cdots ,\mathbf{y}_{k+3}\left( N_b \right) \right]   . \label{eq52}
\end{equation}
According to the assumption of \eqref{eq50}, the block $\mathbf{\dot{y}}_p$ can be restructured into a received block, vectorized block and stacked block via \eqref{eq5}, \eqref{eq10} and \eqref{eq19}, respectively, and be processed by the operations from \eqref{eq35} to \eqref{eq45} to compute the radii of the compressed Gerschgorin circles, denoted by $\left\{ R_{1,p},R_{2,p},\cdots ,R_{J-1,p} \right\}$. Different data blocks transmit over independent channels and have independent data so that we can combine results from different detectors of the data blocks $\mathbf{\dot{y}}$ with different $p$, expressed as 
\begin{equation}
R_{i}=\frac{1}{ {N_d} }\sum_{p=1}^{ {N_d} }{R_{i,p}} \label{eq53}
\end{equation}
where $N_d$ is the number of detectors. Finally, the minimal weighted norm-1 distance metric in \eqref{eq47} decides on the number of transmit antennas and the MIMO scheme using the trained FNNs.

\subsubsection{SFBC-OFDM Case}

Here, we use ${\bf{\dot y}}$ to represent $\mathbf{Y}$, $\mathbf{\bar{y}}$ and $\mathbf{\tilde{y}}$. The estimators of different covariance matrices need to be rewritten as follows
\begin{equation}
\mathbf{\hat{\Sigma}} _{l}=\frac{1}{N_b}\sum_{n=1}^{N_b}{\mathbf{\dot y}_l\left( n \right) \cdot \mathbf{\dot y}_{l}^{H}\left( n \right)} \label{eq54}
\end{equation}
where the subscript $l$ is constrained by the conditions $1\le l\le N$ and has different values for different features as in \eqref{eq34}. We use detectors to calculate the radii of the compressed Gerschgorin circles, $\left\{ R_{1,l},R_{2,l},\cdots ,R_{J-1,l} \right\}$ for each adjacent subcarriers and combine $N_d$ detectors as
\begin{equation}
R_i=\frac{1}{N_d}\sum_{l=1}^{N_d}{R_{i,l}}. \label{eq55}
\end{equation}
Then, the radii are input into the trained FNNs and the features are computed. Finally, the number of transmit antennas and MIMO scheme are determined by the minimal weighted norm-1 distance metric in \eqref{eq47}.

\section{Simulation Results}

\subsection{Simulation Setup}

\subsubsection{Training} 
The training data fed to FNNs are the radii of the compressed Gerschgorin circles and the targets are the NTAF, ICSFs and IRSFs. For the single-carrier system, equiprobable 0/1 data are generated and fed to the transmit diversity encoder after being modulated as $4$-PSK symbols. Then, the symbols are transmitted through MIMO Rayleigh fading channels. After timing and frequency synchronization, the receiver decomposes the received signals and generates the training data via \eqref{eq34}-\eqref{eq45}. For the OFDM system, the differences are additional OFDM operations, frequency-selective fading channels, and generating the final radii via \eqref{eq53}. To achieve a better performance, we retrained the FNNs for the OFDM system since the radii have different distributions between the single-carrier system and OFDM system. This process was repeated 200 times using the Monte Carlo method for each scheme at each SNR. The SNR was varied from -5 dB to 20 dB. The training parameters of the number of receive antennas, observed samples and OFDM symbols, subcarriers, detectors, CP length, modulation type, MIMO scheme pool and channel parameters are as listed in Table \ref{default_sim}.

\subsubsection{Simulation Setting}
\begin{table*}
\centering
\caption{Default system parameters used in the training and simulations.}
\label{default_sim}
\begin{threeparttable}
\begin{tabular}{@{}ccccccccc@{}}
\toprule
System &$N_r$  & $L$/$N_b$  & $N$ & $N_d$  & CP & Modulation & Scheme pool & Channel  \\ 
\midrule
Single-     & \multirow{2}{*}{8}   &  2048\tnote{*} & \multirow{2}{*}{-}   &\multirow{2}{*}{-}  & \multirow{2}{*}{-} & \multirow{2}{*}{4-PSK}     & \multirow{2}{*}{17 types}  & \multirow{2}{*}{Flat-fading with Rayleigh fading coefficients}                                   \\ 
carrier&& 256\tnote{**}&&&&&&    \\
\midrule
\multirow{2}{*}{OFDM}   & \multirow{2}{*}{8}  & 500\tnote{*}   & \multirow{2}{*}{256}   &  64\tnote{*} & \multirow{2}{*}{10} & \multirow{2}{*}{4-PSK}  & \multirow{2}{*}{17 types}  & \multirow{2}{8cm}{Frequency-selective fading consisting of 4 independent complex Gaussian taps with power delay profile $\sigma _t^2 = {e^{ - {t / 5}}}$} \\
&&100\tnote{**} & &32\tnote{**}&&&&\\ \bottomrule
\end{tabular}

 \begin{tablenotes}
        \footnotesize
        \item[*]  Default parameters used in the training.
        \item[**]  Default parameters used in the simulations.
\end{tablenotes}
\end{threeparttable}
\end{table*}

\begin{table}[htbp]
\centering
\caption{Comparison of size of MIMO scheme pool between exiting algorithms and the proposed algorithm.}
\label{pool_comp}
\begin{tabular}{@{}ccc@{}}
\toprule
System                                   & Algorithm            & MIMO scheme pool  \\ \midrule
\multirow{4}{*}{single-carrier}               & \cite{correlator_function,higher_order_cyclic,blind_recognition_STBC,Hierarchical_STBC, Fourth_order_TC,Second_Order_cyclic,K_S_test,Classify_STBC_Over_FS}          &     $\le 5$ types    \\
                                             & \cite{STBC_cyclic_2015_ICC}         &        $ 11$ types           \\
                                             & \cite{Likelihood_Based}          &               $ 13$ types  \\
                                             & Proposed algorithm        &   $ 17$ types   \\ \midrule
\multirow{2}{*}{STBC-OFDM} &  \cite{Blind_MIMO_OFDM,Blind_MIMO_OFDM_SM_AL,Identification_SM_AL_OFDM_cyclic}   &  $\le 3$ types \\
                                             & Proposed algorithm         & $ 17$ types            \\ \midrule
\multirow{2}{*}{SFBC-OFDM} &  \cite{blind_SFBC,My_paper_Globecom,My_paper_TWC2,My_paper_TVT}         &            $\le 5$  types     \\
                                             & Proposed algorithm          & $ 17$ types             \\ \bottomrule
\end{tabular}
\end{table}

Monte Carlo simulations are conducted to evaluate the performance of the proposed algorithm. We consider three systems, namely, the single-carrier, STBC-OFDM, and SFBC-OFDM systems. Unless otherwise mentioned, the default system parameters are as listed in Table \ref{default_sim}. The noise is assumed zero-mean additive white Gaussian with variance $\sigma _n^2$. The total power of transmitted signals is constrained to $P=\left( 1/L \right) \text{E}\left[ \text{Tr}\left( \mathbf{C}\left( \mathbf{x}_b \right) \mathbf{C}^H\left( \mathbf{x}_b \right) \right) \right] $ regardless of the number of transmit antennas $N_t$ and the SNR is defined as $10\log _{10}\left( P/\sigma _{n}^{2} \right) 
$ \cite{Likelihood_Based}. The average probabilities of correct identification ${\rm{Pr}}$ of the number of transmit antennas and MIMO scheme, respectively, are calculated as follows 
\begin{subequations} \label{eq56}
\begin{align}
&{\rm{P}}{{\rm{r}}_1} = \frac{1}{4}\sum {{\rm{Pr}}({{\hat{ N}_t}}|{N_t})} \\
&{\rm{P}}{{\rm{r}}_2} = \frac{1}{{17}}\sum {{\rm{Pr}}({ \hat C}|C)} 
\end{align}  
\end{subequations}
respectively, and used as performance measures. The MIMO scheme pool is defined to contain the 17 types of MIMO schemes listed in Table \ref{table1} (see Appendix C). The simulation of each MIMO scheme was run for 1000 trials.

\subsection{Simulation Results for Single-Carrier System}
We compare the proposed algorithm with the conventional algorithms first, and then evaluate the validity of the proposed algorithm with different system parameters and transmission impairments.

\subsubsection{Performance Evaluation}

First, we compare the size of the MIMO scheme pool between existing algorithms and the proposed algorithm, as shown in Table \ref{pool_comp}. The data on the existing algorithms are based on the survey \cite{Survey_Signal_Identification}. Table \ref{pool_comp} shows that the feature-based algorithms\cite{correlator_function,higher_order_cyclic,blind_recognition_STBC,Hierarchical_STBC, Fourth_order_TC,Second_Order_cyclic,K_S_test,Classify_STBC_Over_FS} are only able to identify less than 5 types of MIMO schemes since many MIMO schemes have the same redundancy locations. The algorithm in \cite{STBC_cyclic_2015_ICC} can identify 11 types of MIMO schemes owing to the pre-processing operation which leads to a finer discriminatory capability. However, this capability depends on \textit{a priori} information including the number of transmit antennas and channel coefficients. The algorithm in \cite{Likelihood_Based} utilizes the code rate to construct a likelihood function which quantifies MIMO schemes with different code rates. There are three approaches introduced in \cite{Likelihood_Based}. The first two approaches require \textit{a priori} information including the number of transmit antennas, channel coefficients and noise power, while the third approach referred to as codes-parameter (COP) based, only requires the number of transmit antennas. The proposed algorithm can identify 17 types of MIMO schemes, and thus has a wider applicability.

\begin{figure}
\centering
\includegraphics[width=0.5\textwidth]{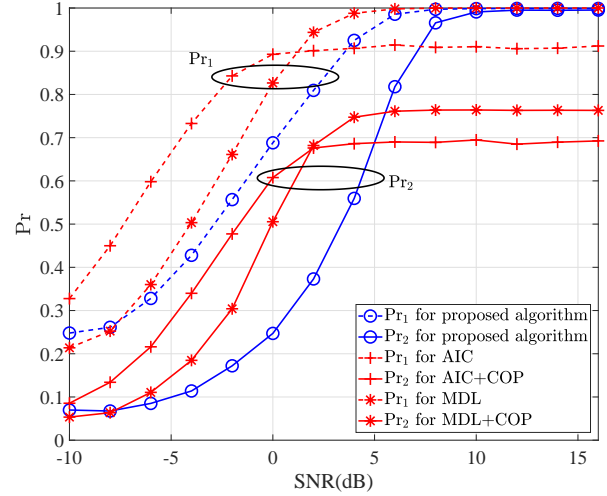}
\caption{Performance comparison between the proposed algorithm and the algorithms AIC/MDL combined with COP based on average probability of correct identification $\rm{Pr}$.}
\label{fig5}
\end{figure}

\begin{table}[htbp]
\centering
\caption{Complexity of the proposed algorithm and comparison.}
\label{complexity}
\begin{tabular}{@{}ccc@{}}
\toprule
\multirow{2}{*}{\centering Algorithm }   & \multirow{2}{2cm}{\centering Computational complexity}  & \multirow{2}{2cm}{\centering Matlab runtime (Average)} \\
                    & &                                                                                                       \\ \midrule
AIC+COP             & $2{\cal{O}}(L+N_{r}^{3})$                                                  & $2.97\  ms$           \\
MDL+COP            & $2{\cal{O}}(L+N_{r}^{3})$                                                 & $3.12\  ms$            \\
Proposed                      & ${\cal{O}}(L+N_{r}^{3})$                                                           & $1.55\  ms$            \\ 
Proposed (OFDM)        & ${\cal{O}}{(N_d}(L+N_{r}^{3}))$  & $31.48\  ms$  \\ \bottomrule
\end{tabular}
\end{table}

For a fair comparison, we compare the combination of AIC/MDL\cite{AIC_MDL} and COP\cite{Likelihood_Based} with the proposed algorithm for the identification of both the number of transmit antennas and MIMO scheme using the same parameters of the single-carrier system described in Table \ref{default_sim}. Fig. \ref{fig5} shows that the algorithms in \cite{AIC_MDL,Likelihood_Based} have a better performance in the low-SNR regime. The reason is that these algorithms use precise mathematical expressions to describe and classify the discriminating features under Rayleigh fading channels which leads to an accurate eigenvalue analysis, while the proposed algorithm employs the heuristic method so that it has a wider applicability. In addition, the probabilities of correct identification of AIC and COP do not converge to one due to the inconsistency of AIC\cite{AIC_MDL} and COP has a smaller pool size (only identifies 13 types).

From a practical point of view, it is important to analyze the computational complexity of the proposed algorithm, which is ${\cal{O}}(L+N_{r}^{3})$, where $N_{r}^{3}$ floating point operations are needed for the eigenvalue decomposition. It is worth noting that this complexity has the same order as those of AIC/MDL or COP, since they require similar operations including the covariance matrix construction and eigenvalue decomposition. We also verify the practical runtime of these algorithms using a computer with a Core i7-7700T CPU, 16 GB RAM (the simulation software is MATLAB$^{\copyright}$ 2017b). The runtime is evaluated using the default parameters listed in Table \ref{default_sim} through 1000 trials. We recorded the runtime for the proposed algorithms, AIC+COP and MDL+COP for each trial and then averaged the runtimes. The average runtime of the proposed algorithm is about 1.55 $ms$, while the combination of AIC/MDL and COP takes about 2.97 $ms$ or 3.12 $ms$. The complexity results are shown in Table \ref{complexity}.

\subsubsection{Effect of the Number of Processed Samples}

\begin{figure}
\centering
\includegraphics[width=0.5\textwidth]{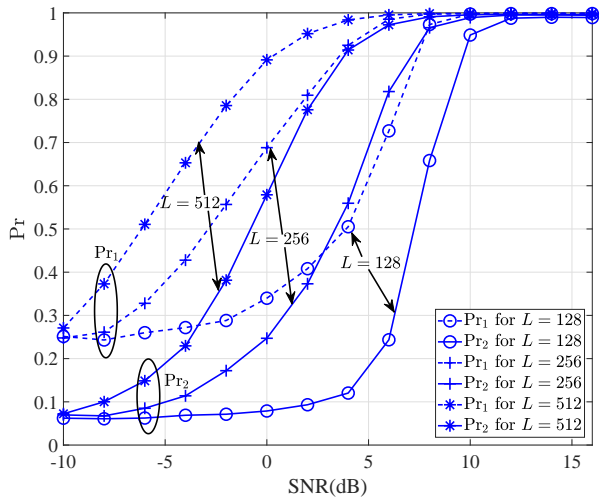}
\caption{Effect of the number of processed samples on the average probability of correct identification $\rm{Pr}$.}
\label{fig6}
\end{figure}

\begin{figure}
\centering
\includegraphics[width=0.5\textwidth]{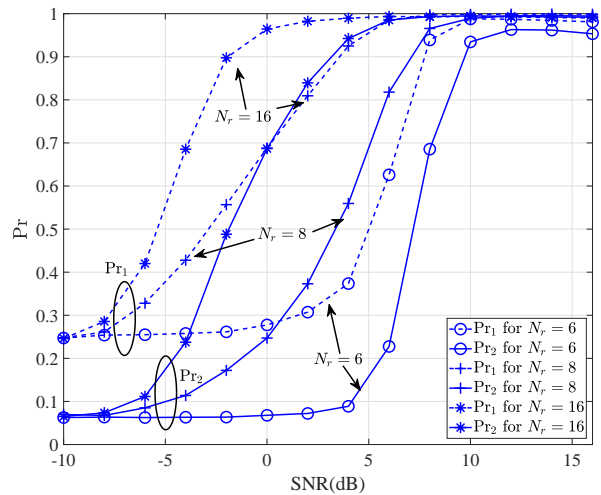}
\caption{Effect of the number of receive antennas on the average probability of correct identification $\rm{Pr}$.}
\label{fig7}
\end{figure}

\begin{figure}
\centering
\includegraphics[width=0.5\textwidth]{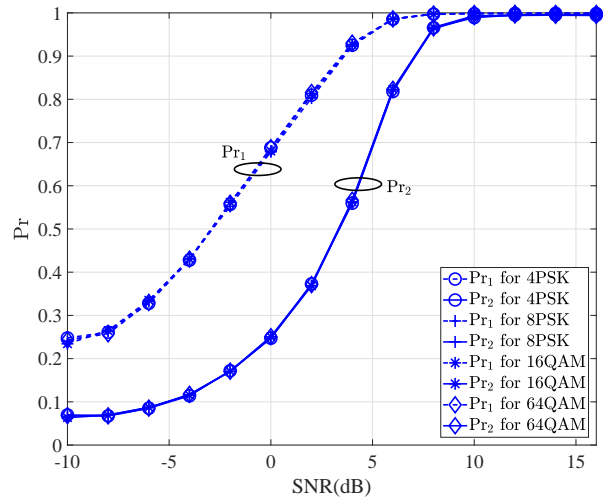}
\caption{Effect of the modulation type on the average probability of correct identification $\rm{Pr}$. }
\label{fig8}
\end{figure}


Fig. \ref{fig6}  shows the performance of the proposed algorithm for different observation intervals. In the three cases, the performance improves with the number of processed samples $L$, because the estimation of the sample covariance matrix in \eqref{eq28} becomes more accurate when $L$ increases.


\subsubsection{Effect of the Number of Receive Antennas}


Fig. \ref{fig7} demonstrates that the probability of correct identification increases with the number of receive antennas for the assumed default simulation parameters. In fact, a large number of receive antennas enhances the estimation performance of the signal subspace dimension, since signals are mapped into a higher dimensional space where it is easy to discriminate between the features. 

\begin{figure}
\centering
\includegraphics[width=0.5\textwidth]{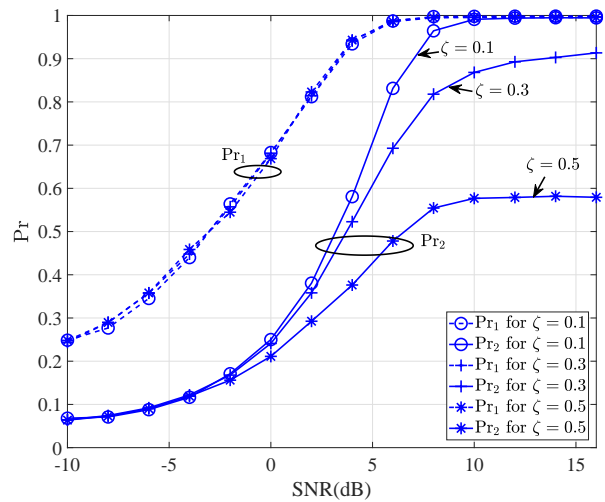}
\caption{Effect of the timing offset on the average probability of correct identification $\rm{Pr}$. }
\label{fig9}
\end{figure}

\begin{figure}
\centering
\includegraphics[width=0.5\textwidth]{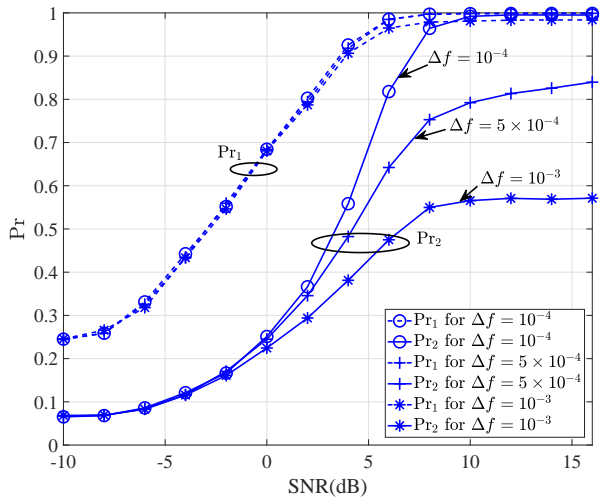}
\caption{Effect of the frequency offset on the average probability of correct identification $\rm{Pr}$. }
\label{fig10}
\end{figure}


\subsubsection{Effect of Modulation Type}
Fig. \ref{fig8} shows the effect of the modulation type on the identification performance. We evaluate the performance with four modulation schemes: 4PSK, 8PSK, 16QAM, 64QAM. These modulations are mandatory for most of the wireless standards. The results demonstrate that the performance does not depend on the modulation type. The reason is that the modulation type does not affect the Gerschgorin circles of the signal subspace since the rank of signal subspace is independent of the modulation type.


\subsubsection{Effect of Timing Offset}

Results presented so far assumed perfect timing synchronization. Now, we evaluate the performance of the proposed algorithm under timing offset. The effect of the timing offset can be modeled as a two-path channel $[1-\zeta, \zeta]$ \cite{time_offset}, where $0 \le \zeta < 1$ is the normalized timing offset when the whole sampling period is one. Fig. \ref{fig9} shows that the timing offset has a small effect on the performance of the identification of the number of transmit antennas, while the effect can be significant on the performance of the identification of MIMO scheme. The reason is that the timing offset destroys the orthogonality of the STBC, which introduces extra terms for the ICSF and IRSF and leads to the overestimation of the features. The timing synchronization parameters can be blindly recovered by algorithms as in \cite{synch_symbol,symbol_recovery} for single-carrier systems and \cite{Cyclic_of_OFDM,OFDM_parameters} for OFDM systems using the cyclostationarity principle.

\begin{figure}
\centering
\includegraphics[width=0.5\textwidth]{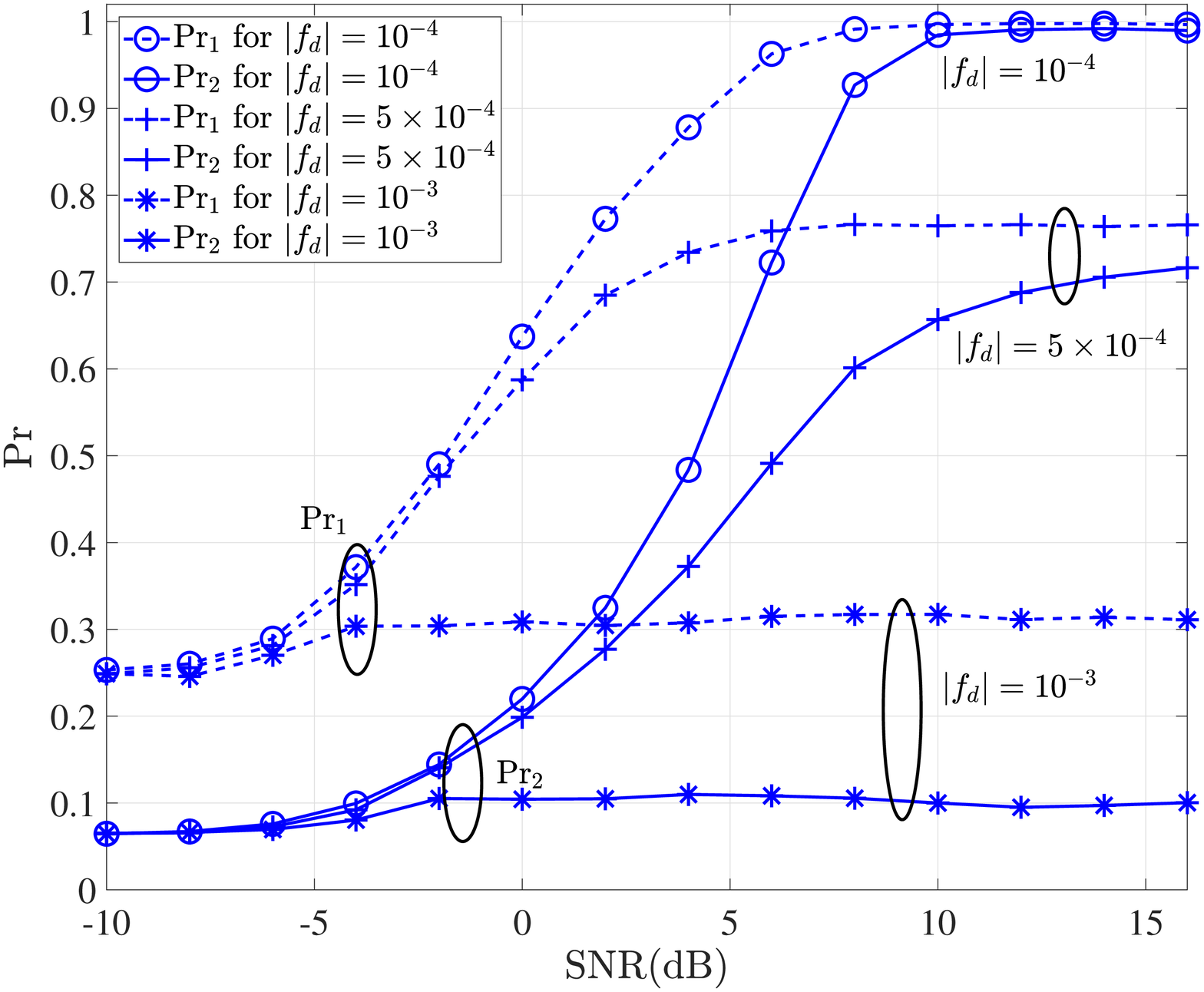}
\caption{Effect of the maximum Doppler frequency on the average probability of correct identification $\rm{Pr}$. }
\label{fig11}
\end{figure}

\begin{figure}
\centering
\includegraphics[width=0.5\textwidth]{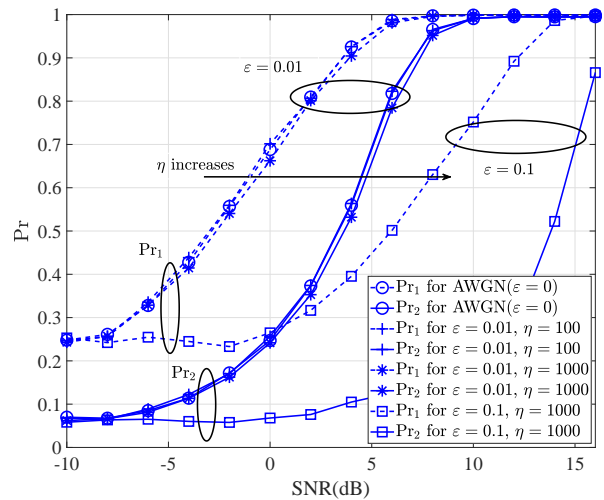}
\caption{Effect of the non-Gaussian noise on the average probability of correct identification $\rm{Pr}$. }
\label{fig14}
\end{figure}


\subsubsection{Effect of Frequency Offset}
Fig. \ref{fig10} illustrates the identification performance of the proposed algorithm for different frequency offsets. The frequency offset, $\Delta f$, is normalized to the signal bandwidth. The identification of the MIMO scheme is impacted by the frequency offset, while the enumeration of the number of transmit antennas is robust with respect to this model mismatch. This is because the frequency offset rotates complex signals and destroys the orthogonality of STBCs, while it does not impact the independence of channels between transmit and receive antennas.  In addition, the frequency offset can also be blindly compensated by an algorithm utilizing the kurtosis-type criterion as in \cite{blind_freq_offset}.

\subsubsection{Effect of Doppler Frequency}
The previous analysis assumed constant channel coefficients over the observation period. Next, we consider the effect of the Doppler spreads on the proposed algorithm. Fig \ref{fig11} shows the identification performance of the proposed algorithm for different Doppler frequencies. Here, the maximum Doppler frequency $|f_d|$ is normalized to the signal bandwidth. The results show a good robustness for $|f_d| < 10^{-4}$. In addition, the Doppler frequency for MIMO signals can also be estimated using a maximum likelihood estimator as in \cite{Blind_Doppler}.


\subsubsection{Effect of Non-Gaussian Noise}
Fig. \ref{fig14} shows the effect of non-Gaussian noise on the proposed algorithm. Here the impulsive noise is modeled as the Gaussian mixture noise with the probability density function (PDF) given by $p(t) = (1- \varepsilon ){\cal N}(0, \sigma^2) + \varepsilon{\cal N}(0, \eta \sigma^2)$,
where $\varepsilon$ is the probability of impulses in noise and ${\cal N}(0, \sigma^2)$ and ${\cal N}(0, \eta \sigma^2)$ denote zero mean Gaussian PDFs with variances $\sigma^2$ and $\eta \sigma^2$, respectively \cite{Non_Gaussian}. The results indicate that the proposed algorithm has a relatively good robustness against the impulsive noise since the Gerschgorin circle-based method can reduce the effect of non-Gaussian noise.

\subsection{Simulation Results for OFDM System}
Our proposed algorithm can identify a larger MIMO scheme pool than existing algorithms, as shown in Table \ref{pool_comp}. In addition, the complexity of the proposed algorithm is ${\cal{O}}{(N_d}(L+N_{r}^{3}))$ due to the use of $N_d$ detectors, as shown in Table \ref{complexity}. 
\subsubsection{STBC-OFDM Case} 
\begin{figure}
\centering
\includegraphics[width=0.5\textwidth]{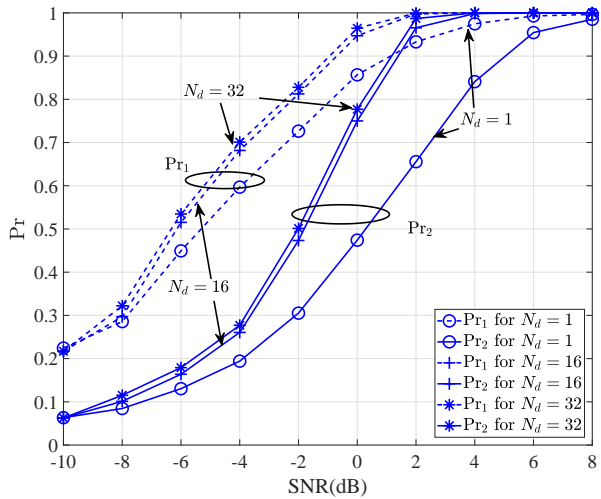}
\caption{Effect of the number of detectors on the average probability of correct identification $\rm{Pr}$ for STBC-OFDM system. }
\label{fig12}
\end{figure}

 Fig. \ref{fig12} demonstrates the viability of the proposed algorithm for STBC-OFDM systems and presents the identification performance for different numbers of detectors, denoted by $N_d$. The performance improves significantly as the number of detectors increases from 1 to 16 with diminishing performance gains beyond 16 detectors. This result indicates that the combination in \eqref{eq49} converges fast with increasing $N_d$. It should also be mentioned that employing one detector makes the proposed algorithm degenerate into the single-carrier system. 

\begin{figure}
\centering
\includegraphics[width=0.5\textwidth]{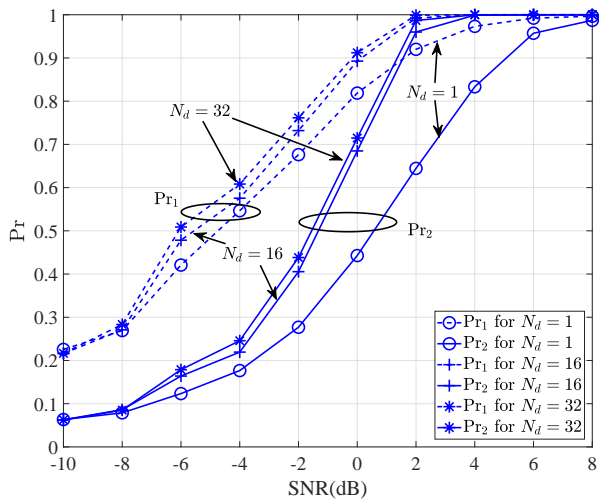}
\caption{Effect of the number of detectors on the average probability of correct identification $\rm{Pr}$ for SFBC-OFDM system.}
\label{fig13}
\end{figure}


\subsubsection{SFBC-OFDM Case}
Fig. \ref{fig13} verifies the viability of the proposed algorithm for the SFBC-OFDM system and illustrates the identification performance for different numbers of detectors. It can be observed that the proposed algorithm for the SFBC-OFDM system has a close performance to that for the STBC-OFDM system. This is because the detector combinations (see Equations \eqref{eq49} and \eqref{eq53}) are the same for these two systems with the same parameters. 

\section{Conclusion and Future Work}

We proposed a novel joint blind identification algorithm for the number of transmit antennas and MIMO schemes. After restructuring the received signals, three  subspace-rank features based on the dimension of the signal subspace, namely, NTAF, ICSF and IRSF, are derived to discriminate between different numbers of transmit antennas and MIMO schemes. Then, we proposed a neural-network Gerschgorin radii-based method to calculate the three features and used a minimal weighted norm-1 distance metric for decision making. Taking advantage of the subspace-rank features and the neural-network Gerschgorin radii-based method, the proposed algorithm can identify a large number of MIMO schemes and applies to both single-carrier and OFDM systems. In addition, the proposed algorithm has an acceptable computational complexity and does not require \textit{a priori} information on the channel coefficients, modulation type or noise power. The simulation results demonstrated the viability of the proposed algorithm for a short observation period both in the single-carrier and OFDM systems, and showed an acceptable performance in the presence of non-Gaussian noise, small timing and frequency offsets and Doppler effects.

The transmission impairments are very challenging problems for the blind identification of MIMO signals and limit their applicability. Future works include devising robust identification algorithms for MIMO signals under relatively large timing and frequency offsets and Doppler effects. Since the analytical expressions of the MIMO signal model fall apart under these impairments, we believe that heuristic approaches are better to address these issues. In addition, deep learning can be a promising approach to the MIMO blind identification problem and we will investigate it in our future work. Furthermore, off-the-air data are planned to be used in future work.

\appendices
\section{TD Examples}

The code matrix of the SM and AL are, respectively, defined as
\begin{align}
& {{\bf{C}}^{{\rm{SM}}}} ( {{{\bf{x}}_b}}  ) = { [ {{x_{b,0}}, \cdots ,{x_{b,{N_t} - 1}}}  ]^T} \notag \\
& {{\bf{C}}^{{\rm{AL}}}} ( {{{\bf{x}}_b}}  ) =  \left[ {\begin{array}{*{20}{c}}
{{x_{b,0}}}&{ - x_{b,1}^ * }\\
{{x_{b,1}}}&{x_{b,0}^ * }
\end{array}}  \right]. \notag
\end{align}

The SBCs using ${N_t} = 3$ transmit antennas are defined by the following coding matrices\cite{STBC_Tarokh,STBC_Ganesan,IEEE802_11}
\begin{equation}
{{\bf{C}}^{{\rm{OSBC}}{{\rm{3}}^1}}}({{\bf{x}}_b}) = {\left[ {\begin{array}{*{20}{c}}
{{x_{b,0}}}&{{x_{b,1}}}&{{x_{b,2}}}\\
{ - {x_{b,1}}}&{{x_{b,0}}}&{ - {x_{b,3}}}\\
{ - {x_{b,2}}}&{{x_{b,3}}}&{{x_{b,0}}}\\
{ - {x_{b,3}}}&{ - {x_{b,2}}}&{{x_{b,1}}}\\
{x_{b,0}^*}&{x_{b,1}^*}&{x_{b,2}^*}\\
{ - x_{b,1}^*}&{x_{b,0}^*}&{ - x_{b,3}^*}\\
{ - x_{b,2}^*}&{x_{b,3}^*}&{x_{b,0}^*}\\
{ - x_{b,3}^*}&{ - x_{b,2}^*}&{x_{b,1}^*}
\end{array}} \right]^T} \notag
\end{equation}
\begin{equation}
{{\bf{C}}^{{\rm{OSBC}}{{\rm{3}}^2}}}({{\bf{x}}_b}) = \left[ {\begin{array}{*{20}{c}}
{{x_{b,0}}}&0&{{x_{b,1}}}&{ - {x_{b,2}}}\\
0&{{x_{b,0}}}&{x_{b,2}^*}&{x_{b,1}^*}\\
{ - x_{b,1}^*}&{ - {x_{b,2}}}&{x_{b,0}^*}&0
\end{array}} \right] \notag
\end{equation}
\begin{equation}
{{\bf{C}}^{{\rm{OSBC}}{{\rm{3}}^3}}}({{\bf{x}}_b}) = \left[ {\begin{array}{*{20}{c}}
{{x_{b,0}}}&{ - x_{b,1}^*}&{x_{b,2}^*}&0\\
{{x_{b,1}}}&{x_{b,0}^*}&0&{ - x_{b,2}^*}\\
{{x_{b,2}}}&0&{ - x_{b,0}^*}&{x_{b,1}^*}
\end{array}} \right] \notag
\end{equation}
\begin{equation}
\mathbf{C}^{{\rm{OSBC3}}^4}\left( \mathbf{x}_b \right) =\left[ \begin{matrix}
	x_{b,0}&		x_{b,1}&		\frac{x_{b,2}}{\sqrt{2}}\\
	-x_{b,1}^{*}&		x_{b,0}^{*}&		\frac{x_{b,2}}{\sqrt{2}}\\
	\frac{x_{b,2}^{*}}{\sqrt{2}}&		\frac{x_{b,2}^{*}}{\sqrt{2}}&		\frac{-x_{b,0}-x_{b,0}^{*}+x_{b,1}-x_{b,1}^{*}}{2}\\
	\frac{x_{b,2}^{*}}{\sqrt{2}}&		-\frac{x_{b,2}^{*}}{\sqrt{2}}&		\frac{x_{b,1}+x_{b,1}^{*}+x_{b,0}-x_{b,0}^{*}}{2}\\
\end{matrix} \right]^T \notag
\end{equation}
\begin{equation}
{{\bf{C}}^{{\rm{SBC3}}}}\left( {{{\bf{x}}_b}} \right) = {\left[ {\begin{array}{*{20}{c}}
{{x_{b,0}}}&{ - x_{b,1}^*}&{{x_{b,2}}}\\
{{x_{b,1}}}&{x_{b,0}^*}&{{x_{b,3}}}
\end{array}} \right]^T} . \notag
\end{equation}


The SBCs and FSTD using ${N_t} = 4$ transmit antennas are defined by the following coding matrices\cite{STBC_Tarokh,STBC_4_ant,QOSTBC,IEEE802_11,sesia2009lte}
\begin{equation}
\mathbf{C}^{\text{OSBC}4^1}\left( \mathbf{x}_b \right) =\left[ \begin{matrix}
	x_{b,0}&		x_{b,1}&		x_{b,2}&		x_{b,3}\\
	-x_{b,1}&		x_{b,0}&		-x_{b,3}&		x_{b,2}\\
	-x_{b,2}&		x_{b,3}&		x_{b,0}&		-x_{b,1}\\
	-x_{b,3}&		-x_{b,2}&		x_{b,1}&		x_{b,0}\\
	x_{b,0}^{*}&		x_{b,1}^{*}&		x_{b,2}^{*}&		x_{b,3}^{*}\\
	-x_{b,1}^{*}&		x_{b,0}^{*}&		-x_{b,3}^{*}&		x_{b,2}^{*}\\
	-x_{b,2}^{*}&		x_{b,3}^{*}&		x_{b,0}^{*}&		-x_{b,1}^{*}\\
	-x_{b,3}^{*}&		-x_{b,2}^{*}&		x_{b,1}^{*}&		x_{b,0}^{*}\\
\end{matrix} \right] ^T \notag
\end{equation}
\begin{figure*}[hb]
  \normalsize
  \vspace*{4pt}
  \hrulefill
  \begin{equation}
\mathbf{C}^{\text{OSBC}4^2}\left( \mathbf{x}_b \right) =\left[ \begin{matrix}
	x_{b,0}&		x_{b,1}&		\frac{x_{b,2}}{\sqrt{2}}&		\frac{x_{b,2}}{\sqrt{2}}\\
	-x_{b,1}^{*}&		x_{b,0}^{*}&		\frac{x_{b,2}}{\sqrt{2}}&		-\frac{x_{b,2}}{\sqrt{2}}\\
	\frac{x_{b,2}^{*}}{\sqrt{2}}&		\frac{x_{b,2}^{*}}{\sqrt{2}}&		\frac{-x_{b,0}-x_{b,0}^{*}+x_{b,1}-x_{b,1}^{*}}{2}&		\frac{x_{b,0}-x_{b,0}^{*}-x_{b,1}-x_{b,1}^{*}}{2}\\
	\frac{x_{b,2}^{*}}{\sqrt{2}}&		-\frac{x_{b,2}^{*}}{\sqrt{2}}&		\frac{x_{b,1}+x_{b,1}^{*}+x_{b,0}-x_{b,0}^{*}}{2}&		\frac{-x_{b,0}-x_{b,0}^{*}-x_{b,1}-x_{b,1}^{*}}{2}\\
\end{matrix} \right] ^T \notag
\end{equation}
  \begin{equation}
\mathbf{C}^{\text{OSBC}4^3}\left( \mathbf{x}_b,\mathbf{x}_{b+1} \right) =\left[ \begin{matrix}
	\mathbf{C}^{\text{AL}}\left( \mathbf{x}_b \right)&		\mathbf{C}^{\text{AL}}\left( \mathbf{x}_{b+1} \right)\\
	-\left[ \mathbf{C}^{\text{AL}}\left( \mathbf{x}_{b+1} \right) \right] ^*&		\frac{\left[ \mathbf{C}^{\text{AL}}\left( \mathbf{x}_{b+1} \right) \right] ^*\mathbf{C}^{\text{AL}}\left( \mathbf{x}_b \right) \mathbf{C}^{\text{AL}}\left( \mathbf{x}_{b+1} \right)}{\lVert \mathbf{x}_{b+1} \rVert ^2}\\
\end{matrix} \right]  \notag
\end{equation}
  \end{figure*}
\begin{equation}
\mathbf{C}^{\text{QOSBC}4}\left( \mathbf{x}_b \right) =\left[ \begin{matrix}
	x_{b,0}&		x_{b,1}&		x_{b,2}&		x_{b,3}\\
	-x_{b,1}^{*}&		x_{b,0}^{*}&		-x_{b,3}^{*}&		x_{b,2}^{*}\\
	-x_{b,2}^{*}&		-x_{b,3}^{*}&		x_{b,0}^{*}&		x_{b,1}^{*}\\
	x_{b,3}&		-x_{b,2}&		-x_{b,1}&		x_{b,0}\\
\end{matrix} \right] \notag
\end{equation}
\begin{equation}
{{\bf{C}}^{{\rm{SBC}}{4^1}}}\left( {{{\bf{x}}_b}} \right) = {\left[ {\begin{array}{*{20}{c}}
{{x_{b,0}}}&{ - x_{b,1}^*}&{{x_{b,2}}}&{ - x_{b,3}^*}\\
{{x_{b,1}}}&{x_{b,0}^*}&{{x_{b,3}}}&{x_{b,2}^*}
\end{array}} \right]^T} \notag
\end{equation}
\begin{equation}
{{\bf{C}}^{{\rm{SBC}}{4^2}}}\left( {{{\bf{x}}_b}} \right) = {\left[ {\begin{array}{*{20}{c}}
{{x_{b,0}}}&{ - x_{b,1}^*}&{{x_{b,2}}}&{{x_{b,4}}}\\
{{x_{b,1}}}&{x_{b,0}^*}&{{x_{b,3}}}&{{x_{b,5}}}
\end{array}} \right]^T} \notag
\end{equation}
\begin{equation}
{{\bf{C}}^{\rm FSTD}}\left( {{{\bf{x}}_b}} \right) = \left[ {\begin{array}{*{20}{c}}
{{x_{b,0}}}&{{x_{b,1}}}&0&0\\
0&0&{{x_{b,2}}}&{{x_{b,3}}}\\
{ - x_{b,1}^*}&{x_{b,0}^*}&0&0\\
0&0&{ - x_{b,3}^*}&{x_{b,2}^*}
\end{array}} \right]. \notag
\end{equation}

\section{Proof of the Propositions}
\subsection{Proof of Proposition 1}
Clearly, the rank of ${{\bf{\Sigma }}_{\bf{S}}}$ is $N_t$, which implies that the rank of the first term on the right hand side of \eqref{eq5} is equal to $N_t$. Thus, all of the smallest ${N_r} - {N_t}$ ordered eigenvalues of ${{\bf{\Sigma }}_{\bf{Y}}}$ are equal to $2 \sigma _w^2$. Q.E.D.

\subsection{Proof of Proposition 2 \& 3}

Assume that an $N \times 1$ random vector $\bf s$ has ${M}$ linearly independent random symbols, and is denoted by ${\bf{ s}}  = { [ {{x_1} , \cdots ,{x_{M}} ,{{x'}_1} , \cdots ,{{x'}_{{N} - {M}}} }  ]^T}$. Then, ${\bf \Sigma} _{\mathbf{s}}$ is given by
\begin{equation}
{\bf \Sigma} _{\mathbf{s}}=\text{E}\left[ \mathbf{ss}^H \right] =\left[ \begin{matrix}
	\text{E}\left[ |x_1|^2 \right]&		\cdots&		\text{E}\left[ x_1\left( x'_{N-M} \right) ^* \right]\\
	\vdots&		\ddots&		\vdots\\
	\text{E}\left[ x_{1}^{*}x'_{N-M} \right]&		\cdots&		\text{E}\left[ |x'_{N-M}|^2 \right]\\
\end{matrix} \right] .
\end{equation}
According to the definition of the linearly independent random symbol, the vector observations ${\bf X'}_1, \cdots, {\bf X'}_{N-M}$ of the random variables ${x'}_1 , \cdots ,{x'}_{N - M}$ can be represented by the linear combination of ${\bf X}_1, \cdots, {\bf X}_{M}$, i.e., ${\bf X'}_i=\sum_{j=1}^{N}{c_j\cdot {\bf X}_j}$ for $i = 1, \cdots, N-M$, where $c_j$ is any real constant. Then, ${\bf \Sigma} _{\mathbf{s}}$ is transformed into a matrix of rank ${M}$ by elementary row operations resulting in the matrix
\begin{equation}
{\bf \Sigma }_{\mathbf{s}}=\left[ \begin{matrix}
	\text{E}\left[ |x_1|^2 \right]&		\cdots&		\text{E}\left[ x_1\left( x'_{N-M} \right) ^* \right]\\
	\vdots&		\ddots&		\vdots\\
	\text{E}\left[ x_{1}^{*}x_M \right]&		\cdots&		\text{E}\left[ x_M\left( x'_{N-M} \right) ^* \right]\\
	\mathbf{0}&		\mathbf{O}&		\mathbf{0}\\
\end{matrix} \right] .
\end{equation}
Assume that $\bf H$ is a $P \times N$ full-rank matrix and $\bf y = Hs+ w$, where $\text{E}\left[ \mathbf{ww}^H \right]= \sigma ^2 {\bf I}_P$. Clearly, all of the smallest $P-M$ ordered eigenvalues of ${{\bf{\Sigma }}_{\bf{y}}} = \text{E}\left[ \mathbf{yy}^H \right]$ are equal to $\sigma ^2$.
Q.E.D.

\section{Subspace-Rank Features for Common MIMO Schemes and a Representative Example}

\begin{table*}[htbp]
\centering
\caption{Features of signals using different numbers of transmit antennas and MIMO schemes.}
\label{table1}
\begin{tabular}{@{}ccccccccc|ccccccccc@{}}
\toprule
\multirow{2}{*}{$N_t$}   &MIMO      &Code     & \multirow{2}{*}{$\alpha$}   &  \multirow{2}{*}{$\beta _1$} &  \multirow{2}{*}{$\beta _2$} &  \multirow{2}{*}{$\beta _3$}  &  \multirow{2}{*}{$\gamma_1$}   &  \multirow{2}{*}{$\gamma_2$}  & \multirow{2}{*}{$N_t$}   &MIMO      &Code     & \multirow{2}{*}{$\alpha$}   &  \multirow{2}{*}{$\beta _1$} &  \multirow{2}{*}{$\beta _2$} &  \multirow{2}{*}{$\beta _3$}  &  \multirow{2}{*}{$\gamma_1$}   &  \multirow{2}{*}{$\gamma_2$}            \\
    &schemes  &rate      & & & & & &    &     &schemes  &rate      & & & & &                               \\
 \midrule
\multirow{2}{*}{1} &Single-                          & \multirow{2}{*}{1}           &  \multirow{2}{*}{1}  & \multirow{2}{*}{2} & \multirow{2}{*}{2}   &  \multirow{2}{*}{2}          &  \multirow{2}{*}{4}   & \multirow{2}{*}{4}   & 3 & ${\rm{SM}}^3$            & 3           & 3  & 6 & 6  & 6         & 12   & 12    \\
 &antenna & & & & & & & &4 & ${\rm{OSBC4}}^1$          & 1/2         & 4  & 4 & 4  & 4        & 8    & 8    \\
2 & $\rm{AL}$                         & 1            & 2 & 4 & 4  & 4          & 4   & 4  & 4 & ${\rm{OSBC4}}^2$          & 3/4         & 4  & 5 & 5  & 5        & 6    & 6   \\
2 & ${\rm{SM}}^2$            & 2           & 2  & 4 & 4  & 4         & 8     & 8  & 4 & ${\rm{OSBC4}}^3$          & 1           & 4  & 8 & 6  & 8         & 8    & 8   \\
3 & ${\rm{OSBC3}}^1$          & 1/2         & 3  & 4 & 4  & 4         & 8  & 8  & 4 & ${\rm{QOSBC}}$              & 1           & 4  & 8 & 4  & 8         & 8    & 8     \\
3 & ${\rm{OSBC3}}^2$          & 3/4         & 3  & 3 & 5  & 5  & 6   & 6 & 4 & ${\rm{FSTD}}$                 & 1           & 4  & 4 & 4  & 4        & 4     & 4        \\
3 & ${\rm{OSBC3}}^3$          & 3/4         & 3 & 5 & 3  & 3   & 6   & 6  & 4 & ${\rm{SBC4}}^1$             & 2           & 4  & 8 & 8  & 8        & 8    & 8   \\
3 & ${\rm{OSBC3}}^4$          & 3/4         & 3 & 5 & 5  & 3   & 6   & 6   & 4 & ${\rm{SBC4}}^2$             & 3           & 4  & 8 & 8  & 8        & 12  & 12    \\
3 & ${\rm{SBC3}}$                & 2           & 3  & 6 & 6  & 6          & 8   & 8  & 4 & ${\rm{SM}}^4$            & 4           & 4  & 8 & 8  & 8        & 16   & 16   \\
\bottomrule
\end{tabular}
\end{table*}

Table \ref{table1} shows all signal features using different numbers of transmit antennas and common MIMO schemes.

\emph{Case of AL:} Assuming $m=1$, we have ${\bf S}(2) = \left[-x_2^*, x_1^*; x_3, x_4 \right]$ ($l=2$). Then, the covariance matrix  ${{\bf{\Sigma }}_{\bf{S}}} (2) = 2{\sigma_s^2}{\bf I}_2$ (assume that the average power of signals is ${\sigma_s^2}$). From {\emph{Proposition 1}}, the rank of the covariance matrix ${{\bf{\Sigma }}_{\bf{Y}}} (2) $ and the cardinality of the set ${\cal A}_2$ are both equal to two. Hence, we have $\alpha = 2$. Subsequently, $\mathbf{\bar{s}}\left( 1 \right) = \left[ x_1, x_2, -x_2^*, x_1^* \right]^T$ ($l_1 = 1$). The covariance matrix ${{\bf{\Sigma }}_{{\bf{\tilde s}}}} (1) = {\sigma_s^2}{\bf I}_4$. From {\emph{Proposition 2}}, the rank of the covariance matrix ${{\bf{\Sigma }}_{{\bf{\tilde y}}}} (1) $ and the cardinality of the set ${\cal B}_1$ are both equal to four. Thus, we have $\beta_1 = 4$. Furthermore, $\mathbf{\tilde{s}}\left( 1 \right)  = \left[ \Re(x_1), \Re(x_2), \Re(-x_2^*), \Re(x_1^*),  \Im(x_1), \Im(x_2), \Im(-x_2^*), \Im(x_1^*) \right]^T$ ($l_1=1$). Then, we have  ${{\bf{\Sigma }}_{{\bf{\tilde s}}}} (1) = \left[ {\sigma _s^2{{\bf{I}}_4}/2, {{\bf{O}}_4}; {{\bf{O}}_4}, {{\bf{O}}_4}} \right]$. Clearly, the rank of the covariance matrix ${{\bf{\Sigma }}_{{\bf{\tilde y}}}} (1)$ and the cardinality of the set ${\cal C}_1$ are both equal to four. Thus, we have $\gamma_1 = 4$.



\ifCLASSOPTIONcaptionsoff
  \newpage
\fi

\bibliographystyle{IEEEtran}

\bibliography{./gao}

\end{document}